\documentclass[10pt,ams]{article}
\usepackage{graphicx}
\usepackage{amssymb}
\usepackage{amsmath}

\newcommand{\p}{\partial}

\newcommand{\s}{\sigma}

\renewcommand{\d}{\delta}
\renewcommand{\S}{\Sigma}
\newcommand{\Int}{\int{d^{4}\!x\,}}

\newcommand{\e}{\varepsilon}

\newcommand{\half}{\mbox{$\frac{1}{2}$}}
\begin{document}
\date{}
\title{\textbf{The gluon and ghost propagators in Euclidean Yang-Mills theory in the maximal Abelian gauge: taking into account
the effects of the Gribov copies and of the dimension two
condensates}}
\author{ \textbf{M.A.L. Capri$^a$\thanks{marcio@dft.if.uerj.br}}\,,
\textbf{V.E.R. Lemes$^{a}$\thanks{vitor@dft.if.uerj.br}}\,,
 \\\textbf{R.F. Sobreiro}$^{b}$\thanks{%
sobreiro@cbpf.br}\;, \textbf{S.P. Sorella}$^{a}$\thanks{%
sorella@uerj.br}\;\footnote{Work supported by FAPERJ, Funda{\c
c}{\~a}o de Amparo {\`a} Pesquisa do Estado do Rio de Janeiro,
under the program {\it Cientista do Nosso Estado},
E-26/100.615/2007.} \,, \textbf{R.
Thibes$^{a}$\thanks{thibes@dft.if.uerj.br}} \\\\
\textit{$^{a}$\small{UERJ
$-$ Universidade do Estado do Rio de Janeiro}}\\
\textit{\small{Instituto de F\'{\i }sica $-$ Departamento de F\'{\i
 }sica Te\'{o}rica}}\\
\textit{\small{Rua S{\~a}o Francisco Xavier 524, 20550-013
Maracan{\~a}, Rio de Janeiro, Brasil}} \\ [2mm]
\textit{$^{b}$\small{CBPF $-$ Centro Brasileiro de Pesquisas
 F\'{\i}sicas}} \\
\textit{\small{Rua Xavier Sigaud 150, 22290-180, Urca, Rio de
Janeiro,
 Brasil}}\\ } \maketitle
\begin{abstract}
The infrared behavior of the gluon and ghost propagators is
studied in $SU(2)$ Euclidean Yang-Mills theory in the maximal
Abelian gauge within the Gribov-Zwanziger framework. The
nonperturbative effects associated with the Gribov copies and with
the dimension two condensates are simultaneously encoded into a
local and renormalizable Lagrangian. The resulting behavior turns
out to be in good agreement with the lattice data.
\end{abstract}

\section{Introduction}
The study of the infrared behavior of the gluon and ghost
propagators has been the object of intensive investigations in
recent years. Albeit not gauge invariant, these correlation
functions enable us to probe the reliability of the various
approaches which give rise to our current understanding of the
behavior of Yang-Mills theories in the infrared, a task which is
far from being achieved. This is due to the fact that propagators
are the simplest Green's functions allowing us to evaluate in
analytic form certain nonperturbative effects expected to be
relevant in the infrared. Moreover, the lattice community has been
able to develop accurate algorithms for a nonperturbative
numerical study of the gluon and ghost propagators, which can be
now analyzed on huge lattices, allowing therefore for a comparison
between analytical and numerical results. Evidently, a qualitative
agreement might be very encouraging in pursuing further
investigations of our theoretical frameworks. We also underline
that this possibility is not restricted to a particular gauge.
Nowadays, the gluon and ghost propagators can be studied from both
theoretical and numerical viewpoints in several gauges as, for
example, the Landau, Coulomb and maximal Abelian gauge.
\\\\In this paper we focus on the gluon and ghost propagators in
the maximal Abelian gauge
\cite{'tHooft:1981ht,Kronfeld:1987vd,Kronfeld:1987ri}, which we
are investigating since several years
\cite{Fazio:2001rm,Lemes:2002ey,Dudal:2002ye,Dudal:2003pe,Dudal:2004rx,Dudal:2005bk,Capri:2005tj,
Gracey:2005vu,Capri:2006vv,Capri:2006cz,Capri:2007hw}. In our
previous works we have provided analytical evidence of
nonperturbative effects which should be taken into account when
facing the various features displayed by this gauge, such as the
dual superconductivity picture for color confinement \cite{scon}
and the Abelian dominance hypothesis
\cite{Ezawa:bf,Suzuki:1989gp,Suzuki:1992gz,Hioki:1991ai}. It turns
out that these nonperturbative effects can be accounted for by a
set of dimension two operators which can be consistently
introduced in the Yang-Mills action.
\\\\As other Lorentz covariant gauges, the maximal Abelian gauge
is plagued by the existence of Gribov copies
\cite{Bruckmann:2000xd}, requiring that the domain of integration
in the Feynman path integral has to be suitably restricted to the
so called Gribov region \cite{Gribov:1977wm}. As discussed in
\cite{Capri:2005tj,Capri:2006cz}, this restriction can be
implemented by introducing a dimension two nonlocal operator,
known as the horizon function, namely
\begin{equation}
S_{\mathrm{hor}}=\gamma^{4}g^{2}\Int \e^{ab}A_{\mu}
\left(\mathcal{M}^{-1}\right)^{ac}\e^{cb}A_{\mu}\;, \label{h}
\end{equation}
where $\gamma$ is the Gribov parameter\footnote{We remind that the
Gribov parameter $\gamma$ is not a free parameter. It is
determined by the gap equation $\frac{\delta \Gamma}{\delta
\gamma}=0$, where $\Gamma$ is the 1PI quantum effective action
evaluated by taking as starting point the Yang-Mills action with
the addition of the horizon term
\cite{Zwanziger:1989mf,Zwanziger:1992qr}. As such, the parameter
$\gamma$ can be expressed in terms of the gauge coupling constant
and of the invariant scale $\Lambda_{QCD}$.} and
$\left(\mathcal{M}^{-1}\right)^{ab}$ is the inverse of
Faddeev-Popov operator
\begin{equation}
\mathcal{M}^{ab}=-D^{ac}_{\mu}D^{cb}_{\mu}
-g^{2}\e^{ac}\e^{bd}A^{c}_{\mu}A^{d}_{\mu}\;, \label{h1}
\end{equation}
with $A_{\mu}$ and $A^{a}_{\mu}$ being the diagonal and
off-diagonal components of the gauge field, respectively, {\it
i.e.} $A_{\mu}=A^{3}_\mu$ and $a=1,2$. Expression \eqref{h}
generalizes to the maximal Abelian gauge the horizon function
already obtained by Zwanziger
\cite{Zwanziger:1989mf,Zwanziger:1992qr} in the Landau gauge. The
nonlocal operator \eqref{h} can be localized by means of the
introduction of a set of auxiliary fields
$(\bar{\phi}^{a}_{i},\phi^{a}_{i}, \bar{\omega}^{a}_{i},
\omega^{a}_{i})$, so that the resulting action enjoys
renormalizability \cite{Capri:2005tj,Capri:2006cz}. \\\\In
addition of the horizon function, other dimension two operators
have been investigated. Our results have given support to the fact
that the preferred vacuum state is that in which those operators
condense, {\it i.e.} they develop a nonvanishing vacuum
expectation value, lowering the vacuum energy of the theory. The
first dimension two operator which has been studied is the gluon
operator $A^{a}_{\mu} A^{a}_{\mu}$. This operator turns out to be
multiplicatively renormalizable \cite{Kondo:2001tm,Dudal:2003pe}
and its condensation, {\it i.e.} $\langle A^{a}_{\mu}
A^{a}_{\mu}\rangle \neq 0$, gives rise to a dynamical mass
generation for off-diagonal gluons \cite{Dudal:2004rx}, in
agreement with the Abelian dominance hypothesis. As second example
of dimension two operator let us quote the ghost operator $\e^{ab}
\bar{c}^{a}c^{b}$, where $\bar{c}^{a}, c^{a}$ denote the
off-diagonal Faddeev-Popov ghosts. This operator is responsible
for the spontaneous breaking of the global $SL(2,\mathbb{R})$
symmetry present in the ghost sector of the maximal Abelian gauge.
It has been investigated by several authors
\cite{Schaden:1999ew,Kondo:2001nq,Lemes:2002ey,Dudal:2002xe}, see
for instance ref.\cite{Capri:2007hw} for a recent analysis of its
renormalizability as well as of its condensation. The third
dimension two operator which we shall consider is given by
$(\bar{\phi}^{a}_{i}\phi^{a}_{i}- \bar{\omega}^{a}_{i}
\omega^{a}_{i}-\bar{c}^a c^a)$. It generalizes to the maximal
Abelian gauge the operator introduced recently in the case of the
Landau gauge \cite{Dudal:2007cw}. It reflects the nontrivial
dynamics developed by the interacting auxiliary fields
$(\bar{\phi}^{a}_{i},\phi^{a}_{i}, \bar{\omega}^{a}_{i},
\omega^{a}_{i})$ needed to localize the horizon term \eqref{h}.
\\\\However, so far, these dimension two operators have not yet
been analyzed simultaneously, a necessary step in order to get a
more precise idea of their relevance in the infrared. This was due
to the nontrivial task of explicitly constructing them, as in the
case of the horizon function, eq.\eqref{h}, as well as to the need
of establishing their renormalizability properties. The aim of
this paper is that of filling this gap, by presenting a detailed
analysis of the gluon and ghost propagators when all these
dimension two operators are present in the starting action. In a
sense, the present work can be seen as a kind of summary of our
efforts towards a better understanding of the infrared behavior of
the gluon and ghost propagators in the maximal Abelian gauge.
\\\\The output of our results can be summarized as follows:
\begin{itemize}
\item when all dimension two operators are simultaneously taken
into account, the resulting local action remains renormalizable.
This nontrivial feature is due to the  large set of Ward
identities which can be established when all operators are
present.

\item the resulting behavior of the gluon and ghost propagators
turns out to be in remarkable agreement with the available lattice
data \cite{Amemiya:1998jz,Bornyakov:2003ee,Mendes:2006kc}. It is
worth underlining that all dimension two operators affect the
propagators. In other words, such a behavior of the gluon and
ghost propagators can be obtained only when the dimension two
operators are simultaneously taken into account.
\end{itemize}
The work is organized as follows. In order to provide a more easy
reading of the paper, in Sect.2 we give a general overview of our
results about the gluon and ghost propagators, providing a
comparison with the recent lattice data. In Sect.3 we present a
detailed discussion of the inclusion in the starting action of the
aforementioned dimension two operators. In Sect.4 we derive the
set of Ward identities fulfilled by the complete action. In Sect.5
we address the issue of the renormalizability of the model. Sect.6
collects our conclusion.

\section{Summary of the results}
\subsection{Notation}
Let us start by briefly reminding the standard notation in the
case of the maximal Abelian gauge. The gauge field ${\cal A}_\mu$
is decomposed as
\begin{equation}
{\cal A}_\mu = A^A_\mu T^A \equiv A^a_{\mu} T^a + A_\mu T^3 \; .
\end{equation}
The generator $T^3$ stands for the diagonal generator of the
$U(1)$ Cartan subalgebra of $SU(2)$, while the index $a =
 1,2 $ labels the remaining off-diagonal generators
 $\{T^a\}$. \\\\Accordingly, the field strength decomposes as
\begin{eqnarray}
 F^a_{\mu\nu} &=& D^{ab}_\mu A^b_\nu - D^{ab}_\nu A^b_\mu
 \;, \nonumber
 \\
 F^3_{\mu\nu} & \equiv & F_{\mu\nu} = \partial_\mu A_\nu - \partial_\nu A_\mu +
 g\e^{ab} A^a_\mu A^b_\nu
 \;, \nonumber
 \\
 \e^{ab} &\equiv & \e^{3ab} \;, \label{not}
\end{eqnarray}
where we have introduced the covariant derivative $D^{ab}_\mu$
with respect to the diagonal components $A_\mu$ of the gauge
field, namely
\begin{equation}
 D^{ab}_\mu \equiv \delta^{ab}\partial_\mu - g\e^{ab} A_\mu
 \label{covdev}
\;.
\end{equation}
\subsection{The tree level gluon and ghost propagators}
We collect here our results for the gluon and ghost propagators.
\begin{itemize}
\item{{\bf  The off-diagonal gluon propagator:} \\the transverse
off-diagonal gluon propagator turns out to be of the Yukawa type
\begin{equation}
\langle A^{a}_{\mu}(-k)A^{b}_{\nu}(k)\rangle = \frac{1}{k^{2}+m^{2}}
\left(\d_{\mu\nu}-\frac{k_{\mu}k_{\nu}}{k^{2}}\right)\d^{ab}\;,
\end{equation}
where $m$ is the dynamical mass originating from the condensation
of the gluon operator \cite{Dudal:2004rx}
\begin{equation}
\mathcal{O}_{A^{2}}=\half A^{a}_{\mu}A^{a}_{\mu}\;.
\label{offgluon}
\end{equation}}
\noindent This behavior has been reported in lattice simulations
\cite{Amemiya:1998jz,Bornyakov:2003ee,Mendes:2006kc}. It supports
the Abelian dominance hypothesis, according to which the
off-diagonal gluons should acquire a sufficiently large
dynamical mass which decouple them at low energies.\\\\
\item{{\bf The diagonal gluon propagator:} \\
for the diagonal gluon propagator we have obtained an infrared
suppressed propagator of the Gribov-Stingl type, namely
\begin{equation}
\langle A_{\mu}(-k)A_{\nu}(k)\rangle =
\frac{k^{2}+\mu^{2}}{k^{4}+\mu^{2}k^{2}+4\gamma^{4}g^{2}}
\left(\d_{\mu\nu}-\frac{k_{\mu}k_{\nu}}{k^{2}}\right)\;,
\label{diaggluon}
\end{equation}
where $\gamma$ is the Gribov parameter and $\mu$ is a mass
parameter related to the condensation of the operator \cite{prep}
\begin{equation}
\mathcal{O}_{\bar{f}f}= (\bar{\phi}^{a}_{i}\phi^{a}_{i} -\bar
\omega^{a}_{i}\omega^{a}_{i} -\bar{c}^{a}c^{a})\;. \label{dp}
\end{equation} }
We observe that expression \eqref{diaggluon} does not vanish at
the origin, in full agreement with the recent numerical data
\cite{Mendes:2006kc}. It gives rise to a positivity violating
propagator in configuration space, a feature usually interpreted
as evidence for gluon confinement.
\\\\Moreover, it is worth to point out that the diagonal gluon
propagator, \eqref{diaggluon}, can be naturally rewritten in terms
of a power-law dynamical running mass of the type
\begin{equation}
\langle A_{\mu}(-k)A_{\nu}(k)\rangle = \frac{1} {k^{2}+M^2(k) }
\left(\d_{\mu\nu}-\frac{k_{\mu}k_{\nu}}{k^{2}}\right)\;,
\label{rm}
\end{equation}
where
\begin{equation}
M^2(k) = \frac{4\gamma^4 g^2}{k^2+\mu^2} \;. \label{M}
\end{equation}
Expression \eqref{rm} is in accordance  with the  definition
firstly envisaged in references
\cite{Cornwall:1981zr,Cornwall:1985bg} and subsequently found  in
the operator product expansion (OPE) approach by
\cite{Lavelle:1991ve}, and later in Schwinger-Dyson equations by
\cite{Aguilar:2007ie}.
\\\\
\item{{\bf The symmetric off-diagonal ghost propagator:}
\\for the symmetric off-diagonal ghost propagator we have found
\begin{equation}
\langle \bar{c}^{a}(-k)c^{b}(k)\rangle_{\rm symm} =
\frac{k^{2}+\mu^{2}}{k^{4}+2\mu^{2}k^{2}
+(\mu^{4}+v^{4})}\,\d^{ab}\;, \label{symmghost}
\end{equation}
where $v$ is a mass parameter related to the condensation of the
ghost operator \cite{Capri:2007hw}
\begin{equation}
\mathcal{O}_{\mathrm{ghost}}= g\e^{ab}\bar{c}^{a}c^{b}\;.
\label{ghop}
\end{equation}}
Notice that expression \eqref{symmghost} is suppressed in the
infrared and attains a nonvanishing finite value at $k=0$. Again,
this behavior agrees with that reported in
\cite{Mendes:2006kc}.\\\\
\item{{\bf The antisymmetric off-diagonal ghost propagator:}
\\finally, for the antisymmetric off-diagonal ghost propagator we
have
\begin{equation}
\langle \bar{c}^{a}(-k)c^{b}(k)\rangle_{\rm antisymm} =
\frac{v^2}{k^{4}+2\mu^{2}k^{2} +(\mu^{4}+v^{4})}\,\e^{ab}\;.
\label{antisymmghost}
\end{equation}}
As expected, this behavior is a consequence of the ghost
condensate \cite{Capri:2007hw},
$\langle\e^{ab}\bar{c}^{a}c^{b}\rangle \sim v^2$, being in
agreement with \cite{Mendes:2006kc}.

\end{itemize}
In summary, the behavior shown above for the gluon and ghost
propagators turns out to be in remarkable agreement with the most
recent lattice data, as reported in \cite{Mendes:2006kc}. This can
be taken as a useful indication of the fact that the
aforementioned dimension two operators play a relevant role in the
infrared. Let also underline that all mass parameters, $(m,
\gamma, \mu, v)$, entering the gluon and ghost propagators are not
free parameters, being determined in a dynamical way as solutions
of gap equations, obtained by minimizing the vacuum energy, see
for instance refs.\cite{Capri:2007hw,Dudal:2004rx} for an estimate
of the values of $m$ and $v$ at one-loop order. As such, all
parameters will get proportional to the unique scale of the
theory, i.e. $(m, \gamma, \mu, v)\propto \Lambda_{QCD}$.

\section{Identification of the complete classical action}
\subsection{The Yang-Mills action and the gauge fixing term}
In order to obtain the complete classical action, let us start by
specifying the gauge fixing term, namely
\begin{equation}
S_{0}=S_{\mathrm{YM}}+S_{\mathrm{MAG}}\;,\label{zero}
\end{equation}
where $S_{\mathrm{YM}}$ is the Yang-Mills action in Euclidean
spacetime
\begin{equation}
S_{\mathrm{YM}}=\frac{1}{4}\Int \Bigl(F^{a}_{\mu\nu}F^{a}_{\mu\nu}
+F_{\mu\nu}F_{\mu\nu}\Bigr)\;,\label{YM}
\end{equation}
with $ F^{a}_{\mu\nu}, F_{\mu\nu}$ and $D^{ab}_{\mu}$ given in
eqs.\eqref{not},\eqref{covdev}.
The term $S_{\mathrm{MAG}}$ in expression \eqref{zero} stands for
the gauge fixing term of the maximal Abelian gauge, being given by
\begin{equation}
S_{\mathrm{MAG}}=\Int\Bigl[\,ib^{a}D^{ab}_{\mu}A^{b}_{\mu}
-\bar{c}^{a}\mathcal{M}^{ab}c^{b}
+g\e^{ab}\bar{c}^{a}(D^{bc}_{\mu}A^{c}_{\mu})c
+ib\,\p_{\mu}A_{\mu} +\bar{c}\,\p_{\mu}(\p_{\mu}c
+g\e^{ab}A^{a}_{\mu}c^{b})\,\Bigr]\;,\label{MAG_action}
\end{equation}
where $(b^{a},b)$ are the off-diagonal and diagonal Lagrange
multipliers enforcing the gauge conditions, given by
$D^{ab}_{\mu}A^{b}_{\mu}=0$ and $\p_{\mu}A_{\mu}=0$. The fields
$(c^{a},\bar{c}^{a},c,\bar{c})$ are the off-diagonal and diagonal
Faddeev-Popov ghosts, respectively, and $\mathcal{M}^{ab}$ denotes
the Faddeev-Popov operator
\begin{equation}
\mathcal{M}^{ab}=-D^{ac}_{\mu}D^{cb}_{\mu}
-g^{2}\e^{ac}\e^{bd}A^{c}_{\mu}A^{d}_{\mu}\;.\label{FP_op}
\end{equation}
The action \eqref{zero} is left invariant by the nilpotent BRST
transformation
\begin{equation}
s^{2}=0\;,
\end{equation}
\begin{equation}
\begin{tabular}{cclccl}
$sA^{a}_{\mu}$&$\!\!\!=\!\!\!$&$-(D^{ab}_{\mu}c^{b}+g\e^{ab}A^{b}_{\mu}c)\,,\qquad$
&$sA_{\mu}$&$\!\!\!=\!\!\!$&$-(\p_{\mu}c+g\e^{ab}A^{a}_{\mu}c^{b})\,,$\vspace{5pt}\\

$sc^{a}$&$\!\!\!=\!\!\!$&$g\e^{ab}c^{b}c\,,$
&$sc$&$\!\!\!=\!\!\!$&$\frac{g}{2}\e^{ab}c^{a}c^{b}\,,$\vspace{5pt}\\

$s\bar{c}^{a}$&$\!\!\!=\!\!\!$&$ib^{a}\,,$
&$s\bar{c}$&$\!\!\!=\!\!\!$&$ib\,,$\vspace{5pt}\\

$sb^{a}$&$\!\!\!=\!\!\!$&$0\,,$&$sb$&$\!\!\!=\!\!\!$&$0\,.$
\end{tabular}
\label{brst_fields}
\end{equation}
Notice that the gauge fixing term \eqref{MAG_action} can be
written as a pure BRST variation
\begin{equation}
S_{\mathrm{MAG}}=s\Int\Bigl(\bar{c}^{a}D^{ab}_{\mu}A^{b}_{\mu}
+\bar{c}\,\p_{\mu}A_{\mu}\Bigr)\;.
\end{equation}
\subsection{Introduction of the horizon function, localization,
and softly broken BRST invariance}
As already mentioned, the maximal Abelian gauge is affected by the
existence of Gribov copies, which have to be taken into account in
order to properly quantize the theory. To deal with this problem
it is necessary to restrict the domain of integration in the
Feynman path integrals to the so-called Gribov region $\Omega$. In
the case of the maximal Abelian gauge, this region is defined as
\cite{Capri:2005tj,Capri:2006cz}
\begin{equation}
\Omega = \{(A^a_{\mu}, A_{\mu}),\;\; D^{ab}_{\mu}A^b_{\mu}=0, \;
\partial_{\mu}A_{\mu}=0, \;\; \mathcal{M}^{ab}=-D^{ac}_{\mu}D^{cb}_{\mu}
-g^{2}\e^{ac}\e^{bd}A^{c}_{\mu}A^{d}_{\mu} > 0 \; \} \;.
\label{gr}
\end{equation}
The restriction of the domain of integration is achieved through
the introduction of the horizon function $S_{\mathrm{hor}}$,
eq.\eqref{h}. Therefore, for the partition function we write
\cite{Capri:2005tj,Capri:2006cz}
\begin{equation}
\mathcal{Z} =\int[dA][db][d\bar{c}][dc]
\,e^{-\left(S_{\mathrm{YM}} +S_{\mathrm{MAG}}
+S_{\mathrm{hor}}\right)}\;.\label{horizon}
\end{equation}
\\\\The nonlocal term
$S_{\mathrm{hor}}$ can be localized by means of a pair of complex
vector bosonic fields, $(\phi^{ab}_{\mu},\bar\phi^{ab}_{\mu})$
according to
\begin{equation}
e^{-S_{\mathrm{hor}}}=\int[d\bar{\phi}][d\phi]\,
\left(\det\mathcal{M}\right)^{8}\, \exp\left\{ -\Int
\left[\,\bar{\phi}^{ab}_{\mu}\mathcal{M}^{ac}\phi^{ab}_{\mu}
+\gamma^{2}g\e^{ab} \left(\phi^{ab}_{\mu}
-\bar{\phi}^{ab}_{\mu}\right)A_{\mu}\,\right] \right\}\;,
\end{equation}
where the determinant, $\left(\det\mathcal{M}\right)^{8}$, takes
into account the Jacobian arising from the integration over the
fields $(\phi^{ab}_{\mu},\bar\phi^{ab}_{\mu})$. This term can also
be localized by means of a pair of complex vector anticommuting
fields $(\omega^{ab}_{\mu},\bar\omega^{ab}_{\mu})$, namely
\begin{equation}
\left(\det\mathcal{M}\right)^{8}=\int [d\bar{\omega}] [d\omega]\,
\exp\left(\,\Int\bar{\omega}^{ab}_{\mu}\mathcal{M}^{ac}
\omega^{cb}_{\mu}\,\right)\;.
\end{equation}
Therefore, the horizon function gives place to a local term
$S_{\mathrm{Local}}$, namely
\begin{eqnarray}
e^{-S_{\mathrm{hor}}}&=&\int[d\bar{\phi}][d\phi][d\bar{\omega}]
[d\omega]\,e^{-S_{\mathrm{Local}}}\;,\nonumber\\
S_{\mathrm{Local}}&=&\Int\left[\,\bar{\phi}^{ab}_{\mu}\mathcal{M}^{ac}
\phi^{cb}_{\mu} -\bar{\omega}^{ab}_{\mu}\mathcal{M}^{ac}
\omega^{cb}_{\mu} +\gamma^{2}g\e^{ab} \left(\phi^{ab}_{\mu}
-\bar{\phi}^{ab}_{\mu}\right)A_{\mu}\,\right]\;.
\end{eqnarray}
Following \cite{Zwanziger:1989mf,Zwanziger:1992qr}, we introduce
the BRST transformations of the localizing fields
$(\phi^{ab}_{\mu},\bar\phi^{ab}_{\mu})$ and
$(\omega^{ab}_{\mu},\bar\omega^{ab}_{\mu})$ as

\begin{equation}
\begin{tabular}{cclccl}
$s\phi^{ab}_{\mu}$&$\!\!\!=\!\!\!$&$\omega^{ab}_{\mu}\;,\qquad$
&$s\omega^{ab}_{\mu}$&$\!\!\!=\!\!\!$&$0\;,\vspace{5pt}$\\

$s\bar{\omega}^{ab}_{\mu}$&$\!\!\!=\!\!\!$&$\bar{\phi}^{ab}_{\mu}\;,$
&$s\bar{\phi}^{ab}_{\mu}$&$\!\!\!=\!\!\!$&$0\;.\vspace{5pt}$\\
\end{tabular}\label{brst_local}
\end{equation}
It should be noted, however, that expression $S_{\mathrm{Local}}$
does not exhibit BRST invariance, which turns out to be broken by
soft terms proportional to the Gribov parameter $\gamma$. In fact
\begin{equation}
s \int d^4x\; \gamma^{2}g \e^{ab} \left(\phi^{ab}_{\mu}
-\bar{\phi}^{ab}_{\mu}\right)A_{\mu} = \gamma^2 g \int d^4x\;
\left[ \e^{ab} \omega^{ab}_{\mu} A_{\mu} - \e^{ab}
\left(\phi^{ab}_{\mu} -\bar{\phi}^{ab}_{\mu}\right) (\partial_\mu
c + g \e^{mn}A^m_{\mu}c^n ) \right] \;. \label{softbr}
\end{equation}
Nevertheless, as in the case of the Landau gauge
\cite{Zwanziger:1989mf,Zwanziger:1992qr,Maggiore:1993wq}, the soft
breaking \eqref{softbr} does not spoil the renormalizability of
the theory \cite{Capri:2005tj,Capri:2006cz}. This remarkable
feature relies on the possibility of extending to the maximal
Abelian gauge the same procedure outlined by Zwanziger in the case
of the Landau gauge \cite{Zwanziger:1989mf,Zwanziger:1992qr},
amounting to embed $S_{\mathrm{Local}}$ into a more general
action, $S_{\mathrm{Local}}^{\mathrm{inv}}$, which enjoys exact
BRST invariance, namely
\begin{equation}
S_{\mathrm{Local}}\to S_{\mathrm{Local}}^{\mathrm{inv}}\;, \qquad
s S_{\mathrm{Local}}^{\mathrm{inv}}= 0 \;.
\end{equation}
Furthermore, as it will be shown below, the term
$S_{\mathrm{Local}}$ can be easily recovered from
$S_{\mathrm{Local}}^{\mathrm{inv}}$. The manifest BRST invariant
action $S_{\mathrm{Local}}^{\mathrm{inv}}$ is found to be
\cite{Capri:2005tj,Capri:2006cz}
\begin{eqnarray}
S_{\mathrm{Local}}^{\mathrm{inv}}&=& s\Int\Bigl(
\bar{\omega}^{ab}_{\mu}\mathcal{M}^{ac}\phi^{cb}_{\mu}
-\bar{N}^{ab}_{\mu\nu}D^{ac}_{\mu}\phi^{cb}_{\nu} +M^{ab}_{\mu
\nu}D^{ac}_{\mu}\bar{\omega}^{cb}_{\nu}\Bigr)\nonumber\\
&=&\Int\Bigl\{\bar{\phi}^{ab}_{\mu}\mathcal{M}^{ac}\phi^{cb}_{\mu}
-\bar{\omega}^{ab}_{\mu}\mathcal{M}^{ac}\omega^{cb}_{\mu}
+\bar{\omega}^{ab}_{\mu}\mathcal{F}^{ac}\phi^{cb}_{\mu}
+\bar{M}^{ab}_{\mu\nu}D^{ac}_{\mu}\phi^{cb}_{\nu} +N^{ab}_{\mu
\nu}D^{ac}_{\mu}\bar{\omega}^{cb}_{\nu}
\nonumber\\
&& +\bar{N}^{ab}_{\mu \nu}[\,D^{ac}_{\mu}\omega^{cb}_{\nu}
+g\e^{ac}(\p_{\mu}c +g\e^{de}A^{d}_{\mu}c^{e})\phi^{cb}_{\nu}\,]
+M^{ab}_{\mu \nu}[\,D^{ac}_{\mu}\bar{\phi}^{cb}_{\nu}
+g\e^{ac}(\p_{\mu}c
+g\e^{de}A^{d}_{\mu}c^{e})\bar{\omega}^{cb}_{\nu}\,]\Bigr\}\;,
\label{S_inv}
\end{eqnarray}
where
\begin{equation}
\mathcal{F}^{ab}=2g\e^{ac}(\p_{\mu}c+g\e^{de}A^{d}_{\mu}c^{e})D^{cb}_{\mu}
+g\e^{ab}\p_{\mu}(\p_{\mu}c+g\e^{cd}A^{c}_{\mu}c^{d})
-g^{2}(\e^{ac}\e^{bd}+\e^{ad}\e^{bc})A^{d}_{\mu}
(D^{ce}_{\mu}c^{e}+g\e^{ce}A^{e}_{\mu}c)\;,
\end{equation}
and the external sources $(M^{ab}_{\mu\nu},\bar{M}^{ab}_{\mu\nu})$,
$(N^{ab}_{\mu\nu},\bar{N}^{ab}_{\mu\nu})$ transform as
\begin{equation}
\begin{tabular}{cclccl}
$sM^{ab}_{\mu\nu}$&$\!\!\!=\!\!\!$&$N^{ab}_{\mu\nu}\,,\qquad$
&$sN^{ab}_{\mu\nu}$&$\!\!\!=\!\!\!$&$0\,,\vspace{5pt}$\\

$s\bar{N}^{ab}_{\mu\nu}$&$\!\!\!=\!\!\!$&$-\bar{M}^{ab}_{\mu\nu}\,,$
&$s\bar{M}^{ab}_{\mu\nu}$&$\!\!\!=\!\!\!$&$0\,.$
\end{tabular}
\end{equation}
In order to reobtain $S_{\mathrm{Local}}$ by the BRST invariant
action $S_{\mathrm{Local}}^{\mathrm{inv}}$ we first take the
physical limits of the external sources
$(M^{ab}_{\mu\nu},\bar{M}^{ab}_{\mu\nu})$,
$(N^{ab}_{\mu\nu},\bar{N}^{ab}_{\mu\nu})$, which are defined by
\cite{Capri:2005tj,Capri:2006cz}
\begin{equation}
\begin{tabular}{cclcl}
$M^{ab}_{\mu\nu}\Bigl|_{\mathrm{phys}}$&$\!\!\!=\!\!\!$&
$-\bar{M}^{ab}_{\mu\nu}\Bigl|_{\mathrm{phys}}$&$\!\!\!=\!\!\!$&
$-\d^{ab}\d_{\mu\nu}\gamma^{2}\;,\vspace{5pt}$\\
$N^{ab}_{\mu\nu}\Bigl|_{\mathrm{phys}}$&$\!\!\!=\!\!\!$&
$-\bar{N}^{ab}_{\mu\nu}\Bigl|_{\mathrm{phys}}$&$\!\!\!=\!\!\!$&$0\;,$
\label{physvalue}
\end{tabular}
\end{equation}
and then perform a shift in the variable $\omega^{ab}_{\mu}$ as
\cite{Capri:2005tj,Capri:2006cz}
\begin{equation}
\omega^{ab}_{\mu}\to\omega^{ab}_{\mu}
+\left(\mathcal{M}^{-1}\right)^{ac}
\left[\,\mathcal{F}^{cd}\phi^{db}_{\mu}
+\gamma^{2}g\e^{cb}(\p_{\mu}c+g\e^{de} A^{d}_{\mu} c^{e})
\right]\;,
\end{equation}
so that
\begin{equation}
S_{\mathrm{Local}}^{\mathrm{inv}}\Bigl|_{\mathrm{phys}} =
S_{\mathrm{Local}}\;.
\end{equation}
Thus, we consider the following action
\begin{equation}
S_{1}=S_{0}+S_{\mathrm{Local}}^{\mathrm{inv}}\;,\label{S1}
\end{equation}
which enjoys the property of being BRST invariant,
\begin{equation}
sS_{1}=0\;.
\end{equation}

\subsection{Inclusion of the quartic ghost term}
Albeit BRST invariant, the action $S_{1}$ is not yet the most
general classical action to start with. The nonlinearity of the
gauge condition, $D^{ab}_{\mu}A^{b}_{\mu}=0$, requires the
introduction of a quartic term in the Faddeev-Popov ghost fields
\begin{equation}
\frac{g^{2}}{2}\,\bar{c}^{a}c^{a} \bar{c}^{b}c^{b}\;,
\end{equation}
which is in fact needed for renormalization purposes. In our case,
due to the presence of the localizing fields $(\phi^{ab}_{\mu},
\bar{\phi}^{ab}_{\mu}, \omega^{ab}_{\mu}, \bar{\omega}^{ab}_{\mu})$,
the quartic ghost term is introduced in a BRST invariant way through
the following invariant action $S_{\alpha}$
\begin{eqnarray}
S_{\alpha}&=&-\frac{\alpha}{2}\,s\Int\left[\,\bar{c}^{a}ib^{a}
-g\e^{ab}\bar{c}^{a}\bar{c}^{b}c +g^{2}\bar{\omega}^{ab}_{\mu}
\phi^{ab}_{\mu}\left(\bar{\phi}^{cd}_{\nu}\phi^{cd}_{\nu}
-\bar{\omega}^{cd}_{\nu}\omega^{cd}_{\nu}\right)
-2g^{2}\bar{\omega}^{ac}_{\mu}\phi^{ac}_{\mu}
\bar{c}^{b}c^{b}\,\right]\nonumber\\
&=&\frac{\alpha}{2}\Int\left[\,b^{a}b^{a}
+2ig\e^{ab}b^{a}\bar{c}^{b}c - g^{2}
\left(\bar{\phi}^{ab}_{\mu}\phi^{ab}_{\mu}
-\bar{\omega}^{ab}_{\mu}\omega^{ab}_{\mu} -\bar{c}^{a}c^{a}\right)
\left(\bar{\phi}^{cd}_{\nu}\phi^{cd}_{\nu}
-\bar{\omega}^{cd}_{\nu}\omega^{cd}_{\nu}
-\bar{c}^{c}c^{c}\right)\right.\nonumber\\
&&\left.-2ig^{2}\bar{\omega}^{ac}_{\mu}\phi^{ac}_{\mu} b^{b}c^{b}
+2g^{3}\bar{\omega}^{ac}_{\mu}\phi^{ac}_{\mu}
\e^{bd}\bar{c}^{b}c^{d}c \,\right]\;,\label{S_alpha}
\end{eqnarray}
where $\alpha$ is a gauge parameter, which has to be set to zero
after the renormalization procedure. In fact, introducing the
action $S_{2}$ as
\begin{equation}
S_{2}=S_{1}+S_{\alpha}\;,
\end{equation}
it follows that the equation of motion of the off-diagonal
Lagrange multiplier $b^{a}$ gets modified according to
\begin{equation}
\frac{\d S_{2}}{\d b^{a}}= iD^{ab}_{\mu}A^{b}_{\mu} +\alpha\left(
b^{a}+ig\e^{ab}\bar{c}^{b}c
-ig^{2}\bar{\omega}^{bc}_{\mu}\phi^{bc}_{\mu}c^{a}\right)\;.
\end{equation}
Therefore, one can see that the gauge condition of the maximal
Abelian gauge, $D^{ab}_{\mu}A^{b}_{\mu}=0$, is attained in the
limit $\alpha\to0$, which has to be taken after the removal of the
ultraviolet divergences. We also remark that the whole term
$S_{\alpha}$ vanishes in the limit $\alpha\to0$, allowing us to
integrate out the localizing fields $(\phi^{ab}_{\mu},
\bar{\phi}^{ab}_{\mu}, \omega^{ab}_{\mu},
\bar{\omega}^{ab}_{\mu})$, and thus recovering the horizon
function \eqref{h}.

\subsection{The global $U(8)$ symmetry}
In addition to the BRST invariance the action $S_{2}$ displays a
global $U(8)$ \cite{Capri:2005tj,Capri:2006cz} symmetry expressed
by
\begin{equation}
\mathcal{Q}^{ab}_{\mu\nu}\,S_{2}=0\;,
\end{equation}
with
\begin{equation}
\mathcal{Q}^{ab}_{\mu\nu}=\Int\!\left(\!\phi^{ca}_{\mu}
\frac{\d}{\d\phi^{cb}_{\nu}}
-\bar{\phi}^{cb}_{\nu}\frac{\d}{\d\bar{\phi}^{ca}_{\mu}}
+\omega^{ca}_{\mu} \frac{\d}{\d\omega^{cb}_{\nu}}
-\bar{\omega}^{cb}_{\nu}\frac{\d}{\d\bar{\omega}^{ca}_{\mu}}
+M^{ca}_{\s\mu}\frac{\d}{\d M^{cb}_{\s\nu}}
-\bar{M}^{cb}_{\s\nu}\frac{\d}{\d\bar{M}^{ca}_{\s\mu}}
+N^{ca}_{\s\mu}\frac{\d}{\d N^{cb}_{\s\nu}}
-\bar{N}^{cb}_{\s\nu}\frac{\d}{\d\bar{N}^{ca}_{\s\mu}}\!\right)\,.
\end{equation}
The presence of the global invariance $U(8)$ means that one can
make use \cite{Capri:2005tj,Capri:2006cz} of the composite index
$i\equiv(a,\mu)$, $i=1,\dots,8$. Therefore, setting
\begin{equation}
(\phi^{ab}_{\mu}, \bar{\phi}^{ab}_{\mu}, \omega^{ab}_{\mu},
\bar{\omega}^{ab}_{\mu}) = (\phi^{a}_{i}, \bar{\phi}^{a}_{i},
\omega^{a}_{i},\bar{\omega}^{a}_{i})\;,
\end{equation}
and
\begin{equation}
(M^{ab}_{\mu\nu},\bar{M}^{ab}_{\mu\nu},
N^{ab}_{\mu\nu},\bar{N}^{ab}_{\mu\nu}) = (M^{a}_{\mu i},
\bar{M}^{a}_{\mu i},N^{a}_{\mu i},\bar{N}^{a}_{\mu i})\;,
\end{equation}
we can write $S_{2}$ as
\begin{eqnarray}
S_{2}&=&S_{\mathrm{YM}}+S_{\mathrm{MAG}}
+\Int\Bigl\{\bar{\phi}^{a}_{i}\mathcal{M}^{ab}\phi^{b}_{i}
-\bar{\omega}^{a}_{i}\mathcal{M}^{ab}\omega^{b}_{i}
+\bar{\omega}^{a}_{i}\mathcal{F}^{ab}\phi^{b}_{i} +\bar{M}^{a}_{\mu
i}D^{ab}_{\mu}\phi^{b}_{i} +N^{a}_{\mu
i}D^{ab}_{\mu}\bar{\omega}^{b}_{i} \nonumber\\
&&+\bar{N}^{a}_{\mu
i}[\,D^{ab}_{\mu}\omega^{b}_{i}+g\e^{ab}(\p_{\mu}c
+g\e^{cd}A^{c}_{\mu}c^{d})\phi^{b}_{i}\,] +M^{a}_{\mu
i}[\,D^{ab}_{\mu}\bar{\phi}^{b}_{i} +g\e^{ab}(\p_{\mu}c
+g\e^{cd}A^{c}_{\mu}c^{d})\bar{\omega}^{b}_{i}\,]\Bigr\}
\nonumber\\
&&+\frac{\alpha}{2}\Int\left[\,b^{a}b^{a}
+2ig\e^{ab}b^{a}\bar{c}^{b}c - g^{2}
\left(\bar{\phi}^{a}_{i}\phi^{a}_{i}
-\bar{\omega}^{a}_{i}\omega^{a}_{i} -\bar{c}^{a}c^{a}\right)
\left(\bar{\phi}^{b}_{j}\phi^{b}_{j}
-\bar{\omega}^{b}_{j}\omega^{b}_{j} -\bar{c}^{b}c^{b}\right)\right.\nonumber\\
&&\left.-2ig^{2}\bar{\omega}^{a}_{i}\phi^{a}_{i} b^{b}c^{b}
+2g^{3}\bar{\omega}^{a}_{i}\phi^{a}_{i} \e^{bc}\bar{c}^{b}c^{c}c
\,\right]\;.\label{S2}
\end{eqnarray}
For the symmetry generator we have
\begin{equation}
\mathcal{Q}_{ij}=\Int\left(\phi^{a}_{i} \frac{\d}{\d\phi^{a}_{j}}
-\bar{\phi}^{a}_{j}\frac{\d}{\d\bar{\phi}^{a}_{i}} +\omega^{a}_{i}
\frac{\d}{\d\omega^{a}_{j}}
-\bar{\omega}^{a}_{j}\frac{\d}{\d\bar{\omega}^{a}_{i}} +M^{a}_{\mu
i}\frac{\d}{\d M^{a}_{\mu j}} -\bar{M}^{a}_{\mu
j}\frac{\d}{\d\bar{M}^{a}_{\mu i}} +N^{a}_{\mu i}\frac{\d}{\d
N^{a}_{\mu j}} -\bar{N}^{a}_{\mu j}\frac{\d}{\d\bar{N}^{a}_{\mu
i}}\right)\;.\label{q8}
\end{equation}
By means of the trace of the operator $\mathcal{Q}_{ij}$,
\textit{i.e.}, $\mathcal{Q}_{ii}\equiv\mathcal{Q}_{8}$, the
$i$-valued fields turn out to possess an additional quantum
number, displayed in the Table \ref{table1}, together with the
mass dimension and the ghost number.

\begin{table}[t]\centering
{\small\begin{tabular}{lcccccccccccccc} \hline\hline &$A$&$b$&$\bar
c$&$c$&$\phi$&$\bar{\phi}$&$\omega$&$\bar{\omega}$&$M$&$\bar{M}$
&$N$&$\bar{N}\!\phantom{\Bigl|}$\\
\hline
dim&$1$&$2$&$2$&$0$&$1$&$1$&$1$&$1$&$2$&$2$&$2$&$2$\\
gh. number&$0$&$0$&$-1$&$1$&$0$&$0$&$1$&$-1$&$0$&$0$&$1$&
$-1$\\
$\mathcal{Q}_{8}$-charge&$0$&$0$&$0$&$0$&$1$&$-1$&$1$&$-1$&$1$&$-1$&$1$&$-1$
\\
\hline\hline
\end{tabular}}
\caption{Quantum numbers of the fields and sources} \label{table1}
\end{table}

\subsection{Introduction of external sources}
In order to establish the set of Ward identities, we have first to
properly define the nonlinear transformations of the fields, as
given in \eqref{brst_fields}. To this purpose, we notice that the
BRST transformation of the gauge field $A^{a}_{\mu}$ can be
written as the sum of two composite operators, namely
\begin{equation}
sA^{a}_{\mu} ={\cal O}_1+{\cal O}_2
\end{equation}
where,
\begin{equation}
{\cal O}_1 = -D^{ab}_{\mu}c^{b},\qquad{\cal
O}_2=-g\varepsilon^{ab}A^{b}_{\mu}c\;.
\end{equation}
Thanks to the fact that the BRST operator is nilpotent, {\it i.e.}
$s^{2}=0$, it follows that
\begin{equation}
s{\cal O}_1=-s{\cal O}_2\;.
\end{equation}
These two operators  can be defined by means of a suitable set of
external sources, $(\Omega^{a}_{\mu},K^{a}_{\mu},\xi^{a}_{\mu})$, as
\begin{equation}
S^{(1)}_{\mathrm{ext}}=\Int
\Bigl[\,\Omega^{a}_{\mu}\left(-D^{ab}_{\mu}c^{b}\right)
+K^{a}_{\mu}\left(-g\varepsilon^{ab}A_{\mu}c\right)
+\xi^{a}_{\mu}\,s\!\left(-g\varepsilon^{ab}A_{\mu}c\right)\Bigr]\;.
\end{equation}
To guarantee the BRST invariance of $S^{(1)}_{\mathrm{ext}}$ we
require that
\begin{equation}
s\xi^{a}_{\mu}=-(\Omega^{a}_{\mu}-K^{a}_{\mu})\,,\qquad
s\Omega^{a}_{\mu}=sK^{a}_{\mu}=0\;.
\end{equation}
The nonlinear BRST transformations of the fields $A_{\mu}, c^{a},
c$ can be accounted for by the external sources $\Omega_{\mu},
L^a, L$, according to
\begin{eqnarray}
S^{(2)}_{\mathrm{ext}}&=& s\Int\Bigl[-\Omega_{\mu}A_{\mu}
+L^{a}c^{a} +Lc \Bigr]\nonumber\\
&=&\Int\Bigl[-\Omega_{\mu}(\p_{\mu}c +g\e^{ab}A^{a}_{\mu}c^{b})
+g\e^{ab}L^{a}c^{b}c +\frac{g}{2}\e^{ab}Lc^{a}c^{b}\Bigr]\;,
\end{eqnarray}
where we require that
\begin{equation}
s\Omega_{\mu}=0\,,\qquad sL^{a}=0\,,\qquad sL=0\;.
\end{equation}
Moreover, adding $S^{(1)}_{\mathrm{ext}}$ to $S_{2}$ we obtain an
action that is left invariant by the following transformations:
\begin{equation}
\begin{tabular}{cclccl}
\multicolumn{3}{l}{\textbf{The $\d_{i}$ symmetry:}}
&\multicolumn{3}{l}{\textbf{The $\bar\d_{i}$ symmetry:}}\vspace{5pt}\\
$\d_{i}\bar{c}^{a}$&$\!\!\!=\!\!\!$&$\phi^{a}_{i}\,,\qquad\qquad\qquad$
&$\bar{\d}_{i}\bar{c}^{a}$&$\!\!\!=\!\!\!$&$\bar{\omega}^{a}_{i}\,,
\vspace{5pt}$\\

$\d_{i}\bar{\phi}^{a}_{j}$&$\!\!\!=\!\!\!$&$\d_{ij}\,c^{a}\,,$
&$\bar{\d}_{i}\omega^{a}_{j}$&$\!\!\!=\!\!\!$&$-\d_{ij}\,c^{a}\,,\vspace{5pt}$\\

$\d_{i}b^{a}$&$\!\!\!=\!\!\!$&$-ig\e^{ab}\phi^{b}_{i}c\,,$
&$\bar{\d}_{i}b^{a}$&$\!\!\!=\!\!\!$&$-ig\e^{ab}\bar{\omega}^{b}_{i} c\,,\vspace{5pt}$\\

$\d_{i}\Omega^{a}_{\mu}$&$\!\!\!=\!\!\!$&$M^{a}_{\mu i}\,,$
&$\bar{\d}_{i}\Omega^{a}_{\mu}$&$\!\!\!=\!\!\!$&$-\bar{N}^{a}_{\mu
i}\,.$
\end{tabular}\label{symmetries}
\end{equation}
As transformations \eqref{symmetries} contain composite field
operators, \textit{i.e.}, $g\e^{ab}\phi^{b}_{i}c$ and
$g\e^{ab}\bar{\omega}^{b}c$, we define them by means of additional
external sources $(Y^{a}_{i},X^{a}_{i})$ and
$(\bar{X}^{a}_{i},\bar{Y}^{a}_{i})$, giving rise to two sets of
BRST doublets
\begin{equation}
\begin{tabular}{cclccl}
$sY^{a}_{i}$&$\!\!\!=\!\!\!$&$X^{a}_{i}\,,\qquad$
&$sX^{a}_{i}$&$\!\!\!=\!\!\!$&$0\,,\vspace{5pt}$\\

$s\bar{X}^{a}_{i}$&$\!\!\!=\!\!\!$&$-\bar{Y}^{a}_{i}\,,$
&$s\bar{Y}^{a}_{i}$&$\!\!\!=\!\!\!$&$0\,,$
\end{tabular}
\end{equation}
so that
\begin{eqnarray}
S^{(3)}_{\mathrm{ext}}&=&-s\Int
g\e^{ab}\left(\bar{X}^{a}_{i}\phi^{b}_{i}c
-Y^{a}_{i}\bar{\omega}^{b}_{i}c\right)\nonumber\\
&=&\Int\Bigl[g\e^{ab}\bar{Y}^{a}_{i}\phi^{b}_{i}c
-\bar{X}^{a}_{i}\Bigl(g\e^{ab}\omega^{b}_{i}c
+\frac{g^{2}}{2}\e^{ab}\e^{cd}\phi^{b}_{i}c^{c}c^{d}\Bigr)\nonumber\\
&&+g\e^{ab}X^{a}_{i}\bar{\omega}^{b}_{i}c
-Y^{a}_{i}\Bigl(g\e^{ab}\bar{\phi}^{b}_{i}c
-\frac{g^{2}}{2}\e^{ab}\e^{cd}\bar{\omega}^{b}_{i}c^{c}c^{d}\Bigl)\Bigr]\;.
\end{eqnarray}
Therefore, for the most general invariant external source term
which can be added to $S_{2}$, we obtain
\begin{equation}
S_{\mathrm{ext}}=S_{\mathrm{ext}}^{(1)} +S_{\mathrm{ext}}^{(2)}
+S_{\mathrm{ext}}^{(3)} +\chi\Int\left( \bar{M}^{a}_{\mu
i}M^{a}_{\mu i} +\bar{N}^{a}_{\mu i}N^{a}_{\mu
i}\right)\;,\label{ext}
\end{equation}
where the last term, which can be written as an exact BRST
variation
\begin{equation}
\chi\Int\left( \bar{M}^{a}_{\mu i}M^{a}_{\mu i} +\bar{N}^{a}_{\mu
i}N^{a}_{\mu i}\right)=-\chi s\Int\bar{N}^{a}_{\mu i}M^{a}_{\mu
i}\;,
\end{equation}
is allowed by power counting and has to be added for
renormalization purposes. Also, the parameter $\chi$ stands for a
free coefficient.

\subsection{Introduction of dimension two operators}
The last step towards the construction of the complete starting
classical action is the introduction of the three dimension two
operators $\mathcal{O}_{A^{2}}, \mathcal{O}_{\bar{f}f}$ and
$\mathcal{O}_{\mathrm{ghost}}$,
eqs.\eqref{offgluon},\eqref{dp},\eqref{ghop}. Let us start by
considering the gluon operator
$\mathcal{O}_{A^{2}}(x)=\half\,A^{a}_{\mu}(x)A^{a}_{\mu}(x)$.
Introducing the BRST doublet of sources $(\lambda,J)$ as
\begin{equation}
s\lambda=J\;,\qquad sJ=0\;,
\end{equation}
it turns out that $\mathcal{O}_{A^{2}}$ can be introduced in a
BRST invariant way, namely
\begin{equation}
S_{J}=s\Int\lambda\left(\mathcal{O}_{A^{2}} +\half\zeta\,J\right)
=\Int\left(J\mathcal{O}_{A^{2}}+\half\zeta\,J^{2}
+\lambda\,A^{a}_{\mu} D^{ab}_{\mu}c^b\right)\;,\label{Sj}
\end{equation}
where $\zeta$ is a constant parameter needed to account for the
ultraviolet divergences of the vacuum correlation function
$\langle
(A^{a}_{\mu}(x)A^{a}_{\mu}(x))(A^{b}_{\nu}(y)A^{b}_{\nu}(y))
\rangle$.
\\\\The operators $\mathcal{O}_{\bar{f}f}, \mathcal{O}_{\mathrm{ghost}}$
can be introduced in a similar way. More specifically, defining
the BRST doublet of sources $(\tau,\s)$ as
\begin{equation}
s\tau=\s\;,\qquad s\s=0\;,
\end{equation}
the invariant term $S_{\s}$ describing the coupling of
$\mathcal{O}_{\bar{f}f}$ is given by
\begin{equation}
S_{\s}=s\Int\tau\left(\mathcal{O}_{\bar{f}f} +\half\kappa\,\s
+\rho\,J\right)
=\Int\left(\s\mathcal{O}_{\bar{f}f}+\half\kappa\,\s^{2} +\rho\,\s
J -\tau\,s\mathcal{O}_{\bar{f}f}\right)\;,\label{S_sigma}
\end{equation}
where $\kappa$ and $\rho$ are constant parameters, needed for
renormalization purposes. Notice in fact that expression
\eqref{S_sigma} contains the mixing term $\s J$. This term,
allowed by power counting, accounts for the ultraviolet
divergences of the mixed vacuum correlation function $\langle
(A^{a}_{\mu}(x)A^{a}_{\mu}(x))
(\bar{\phi}^{b}_{i}(y)\phi^{b}_{i}(y)
-\bar{\omega}^{b}_{i}(y)\omega^{b}_{i}(y) -\bar{c}^{b}(y)c^{b}(y))
\rangle$. \\\\Finally, the introduction of a third doublet of
sources $(\eta,\theta)$
\begin{equation}
s\eta=\theta\;,\qquad s\theta=0\;,
\end{equation}
allows us to introduce the ghost operator
$\mathcal{O}_{\mathrm{ghost}}(x)=g\e^{ab}\bar{c}^{a}(x)c^{b}(x)$,
namely
\begin{equation}
S_{\theta}=s\Int \eta\left(\mathcal{O}_{\mathrm{ghost}}
+\half\beta\,\theta \right)
=\Int\left(\theta\mathcal{O}_{\mathrm{ghost}}+\half\beta\,\theta^{2}
-\eta\,s\mathcal{O}_{\mathrm{ghost}}\right)\;,\label{S_theta}
\end{equation}
where $\beta$ is a constant parameter needed for the divergences
of the correlation function
$\langle(\e^{ab}\bar{c}^{a}(x)c^{b}(x))(\e^{mn}\bar{c}^{m}(y)c^{n}(y))\rangle$.
Notice, however, that the ghost operator
$\mathcal{O}_{\mathrm{ghost}}$ breaks the symmetries
\eqref{symmetries}. Therefore, to maintain the symmetry content of
the theory is necessary to introduce two more BRST doubles of
external sources,
\begin{equation}
\begin{tabular}{cclccl}
$s\eta_{i}$&$\!\!\!=\!\!\!$&$-\theta_{i}\,,\qquad$
&$s\theta_{i}$&$\!\!\!=\!\!\!$&$0\,,\vspace{5pt}$\\

$s\bar{\theta}_{i}$&$\!\!\!=\!\!\!$&$\bar{\eta}_{i}\,,$
&$s\bar{\eta}_{i}$&$\!\!\!=\!\!\!$&$0\,,$
\end{tabular}
\end{equation}
and define an extra term given by
\begin{eqnarray}
S_{\mathrm{extra}}&=&s\Int
g\e^{ab}\left(\bar{\theta}_{i}\phi^{a}_{i}c^{b}
-\eta_{i}\bar{\omega}^{a}_{i}c^{b}\right)\nonumber\\
&=&\Int \left[\,g\e^{ab}\left(\bar{\eta}_{i}\phi^{a}_{i}c^{b}
+\eta_{i}\bar{\phi}^{a}_{i}c^{b}
+\bar{\theta}_{i}\omega^{a}_{i}c^{b}
+\theta_{i}\bar{\omega}^{a}_{i}c^{b}\right)
-g^{2}\bar{\theta}_{i}\phi^{a}_{i}c^{a}c
+g^{2}\eta_{i}\bar{\omega}^{a}_{i}c^{a}c\,\right]\;.\label{extra}
\end{eqnarray}

\subsection{The complete classical action}
We are now ready to write down the complete classical action $\S$,
given by
\begin{eqnarray}
\S&=&\S_{0}+S_{\mathrm{extra}}\;,\nonumber\\
\S_{0}&=&S_{2}+S_{\mathrm{ext}}+S_{J}+S_{\s}+S_{\theta}\;,\nonumber\\
S_{2}&=&S_{1}+S_{\alpha}\;,\nonumber\\
S_{1}&=&S_{0}+S_{\mathrm{Local}}^{\mathrm{inv}}\;,\nonumber\\
S_{0}&=&S_{\mathrm{YM}} +S_{\mathrm{MAG}}\;,
\end{eqnarray}
where, $S_{\mathrm{YM}}$, $S_{\mathrm{MAG}}$,
$S_{\mathrm{Local}}^{\mathrm{inv}}$, $S_{\alpha}$,
$S_{\mathrm{ext}}$, $S_{J}$, $S_{\s}$, $S_{\theta}$,
$S_{\mathrm{extra}}$ are given, respectively, by \eqref{YM},
\eqref{MAG_action}, \eqref{S_inv}, \eqref{S_alpha}, \eqref{ext},
\eqref{Sj}, \eqref{S_sigma}, \eqref{S_theta}, \eqref{extra}. Thus,
the complete classical action is
\begin{equation}
\S=S_{\mathrm{YM}} +S_{\mathrm{MAG}}
+S_{\mathrm{Local}}^{\mathrm{inv}} +S_{\alpha} +S_{\mathrm{ext}}
+S_{J} +S_{\s} +S_{\theta} +S_{\mathrm{extra}}\;,
\end{equation}
or, explicitly, we have
\begin{eqnarray}
\Sigma&=&S_{\mathrm{YM}}
+s\Int\Bigl(\bar{c}^{a}D^{ab}_{\mu}A^{b}_{\mu}
+\bar{c}\,\p_{\mu}A_{\mu}+\bar{\omega}^{a}_{i}\mathcal{M}^{ab}\phi^{b}_{i}
-\bar{N}^{a}_{\mu i}D^{ab}_{\mu}\phi^{b}_{i}
+M^{a}_{\mu i}D^{ab}_{\mu}\bar{\omega}^{b}_{i}\nonumber\\
&&-g\e^{ab}\bar{X}^{a}_{i}\phi^{b}_{i}c
+g\e^{ab}Y^{a}_{i}\bar{\omega}^{b}_{i}c -\Omega^{a}_{\mu}A^{a}_{\mu}
-g\e^{ab}\xi^{a}_{\mu}A^{b}_{\mu}c - \Omega_{\mu}A_{\mu} +L^{a}c^{a}
+Lc
-\chi\bar{N}^{a}_{\mu i}M^{a}_{\mu i}\Bigl)\nonumber\\
&&-\frac{\alpha}{2}s\Int\Bigl[\,\bar{c}^{a}ib^{a}
-g\e^{ab}\bar{c}^{a}\bar{c}^{b}c
+g^{2}\bar{\omega}^{a}_{i}\phi^{a}_{i}(\bar{\phi}^{b}_{j}\phi^{b}_{j}
-\bar{\omega}^{b}_{j}\omega^{b}_{j})
-2g^{2}\bar{\omega}^{a}_{i}\phi^{a}_{i}
\bar{c}^{b}c^{b}\,\Bigl]\nonumber\\
&&+s\Int\lambda\left(\half A^{a}_{\mu}A^{a}_{\mu}+\half\zeta\,
J\right) +s\Int\tau\left[\left(\bar{\phi}^{a}_{i}\phi^{a}_{i}
-\bar{\omega}^{a}_{i}\omega^{a}_{i}
-\bar{c}^{a}c^{a}\right)+\half\kappa\,\s +\rho\,J\,\right]\nonumber\\
&&+s\Int \left[g\e^{ab}\left(\eta\bar{c}^{a}c^{b}
+\bar{\theta}_{i}\phi^{a}_{i}c^{b}
-\eta_{i}\bar{\omega}^{a}_{i}c^{b}\right)
+\half\beta\,\eta\theta\right]\nonumber\cr
&=&S_{\mathrm{YM}}+\Int\Bigl\{ib^{a}D^{ab}_{\mu}A^{b}_{\mu}
-\bar{c}^{a}\mathcal{M}^{ab}c^{b}
+g\e^{ab}\bar{c}^{a}(D^{bc}_{\mu}A^{c}_{\mu})c
+ib\,\p_{\mu}A_{\mu}
+\bar{c}\,\p_{\mu}(\p_{\mu}c+g\e^{ab}A^{a}_{\mu}c^{b})
+\bar{\phi}^{a}_{i}\mathcal{M}^{ab}\phi^{b}_{i}\nonumber\\
&&-\bar{\omega}^{a}_{i}\mathcal{M}^{ab}\omega^{b}_{i}
+\bar{\omega}^{a}_{i}\mathcal{F}^{ab}\phi^{b}_{i} +\bar{M}^{a}_{\mu
i}D^{ab}_{\mu}\phi^{b}_{i} +\bar{N}^{a}_{\mu
i}[\,D^{ab}_{\mu}\omega^{b}_{i}
+g\e^{ab}(\p_{\mu}c+g\e^{cd}A^{c}_{\mu}c^{d})\phi^{b}_{i}\,]
+N^{a}_{\mu i}D^{ab}_{\mu}\bar{\omega}^{b}_{i}\nonumber\\
&&+M^{a}_{\mu i}[\,D^{ab}_{\mu}\bar{\phi}^{b}_{i}
+g\e^{ab}(\p_{\mu}c+g\e^{cd}A^{c}_{\mu}c^{d})\bar{\omega}^{b}_{i}\,]
-\Omega^{a}_{\mu}D^{ab}_{\mu}c^{b}
-g\e^{ab}K^{a}_{\mu}A^{b}_{\mu}c
+\xi^{a}_{\mu}\Bigl[\,g\e^{ab}(D^{bc}_{\mu}c^{c})c
\nonumber\\
&&-\frac{g^{2}}{2}\e^{ab}\e^{cd}A^{b}_{\mu}c^{c}c^{d}\,\Bigr]
-\Omega_{\mu}(\p_{\mu}c +g\e^{ab}A^{a}_{\mu}c^{b})
+g\e^{ab}L^{a}c^{b}c+\frac{g}{2}\e^{ab}Lc^{a}c^{b}
+g\e^{ab}\bar{Y}^{a}_{i}\phi^{b}_{i}c
-\bar{X}^{a}_{i}\Bigl[g\e^{ab}\omega^{b}_{i}c\nonumber\\
&&+\frac{g^{2}}{2}\e^{ab}\e^{cd}\phi^{b}_{i}c^{c}c^{d}\,\Bigr]
+g\e^{ab}X^{a}_{i}\bar{\omega}^{b}_{i}c
-Y^{a}_{i}\Bigl[g\e^{ab}\bar{\phi}^{b}_{i}c
-\frac{g^{2}}{2}\e^{ab}\e^{cd}\bar{\omega}^{b}_{i}c^{c}c^{d}\Bigr]
+\chi\left(\bar{M}^{a}_{\mu i}M^{a}_{\mu i} +\bar{N}^{a}_{\mu
i}N^{a}_{\mu i}\right) \nonumber\\
&&+\frac{\alpha}{2}[\,b^{a}b^{a} +2ig\e^{ab}b^{a}\bar{c}^{b}c
-g^{2}\left(\bar{\phi}^{a}_{i}\phi^{a}_{i}
-\bar{\omega}^{a}_{i}\omega^{a}_{i} -\bar{c}^{a}c^{a}\right)
\left(\bar{\phi}^{b}_{j}\phi^{b}_{j}
-\bar{\omega}^{b}_{j}\omega^{b}_{j} -\bar{c}^{b}c^{b}\right)
-2ig^{2}\bar{\omega}^{a}_{i}\phi^{a}_{i}\,b^{b}c^{b}
\nonumber\\
&&+2g^{3}\bar{\omega}^{a}_{i}\phi^{a}_{i}
\,\varepsilon^{bc}\bar{c}^{b}c^{c}c\,]+\half
J\,A^{a}_{\mu}A^{a}_{\mu}+\lambda\,A^{a}_{\mu}D^{ab}_{\mu}c^{b}
+\s\left(\bar{\phi}^{a}_{i}\phi^{a}_{i}
-\bar{\omega}^{a}_{i}\omega^{a}_{i} -\bar{c}^{a}c^{a}\right)
+\tau\left(ib^{a}c^{a} -g\e^{ab}\bar{c}^{a}c^{b}c\right)\nonumber\\
&&+\frac{\zeta}{2}J^{2}+\rho\,J\s+\frac{\kappa}{2}\sigma^{2}
+g\e^{ab}\left(\theta\bar{c}^{a}c^{b}+\bar{\eta}_{i}\phi^{a}_{i}c^{b}
+\eta_{i}\bar{\phi}^{a}_{i}c^{b}
+\bar{\theta}_{i}\omega^{a}_{i}c^{b}
+\theta_{i}\bar{\omega}^{a}_{i}c^{b}\right)
-ig\e^{ab}\eta b^{a}c^{b}-g^{2}\eta\bar{c}^{a}c^{a}c\nonumber\\
&&-g^{2}\bar{\theta}_{i}\phi^{a}_{i}c^{a}c
+g^{2}\eta_{i}\bar{\omega}^{a}_{i}c^{a}c
+\frac{\beta}{2}\,\theta^{2}\Bigr\}\;. \label{clact}
\end{eqnarray}
The expressions for the gluon and ghost propagators given in
Sect.2 are easily derived by considering the relevant quadratic
terms of \eqref{clact} and by replacing $(J,\s,\theta)$ by the
more conventional mass parameters $(m^2, \mu^2, v^2)$ originating
from the corresponding dimension two condensates, {\it i.e.} $m^2
\sim \langle A^a_{\mu} A^a_{\mu} \rangle$ \cite{Dudal:2004rx}, $
\mu^2 \sim \langle\left(\bar{\phi}^{a}_{i}\phi^{a}_{i}
-\bar{\omega}^{a}_{i}\omega^{a}_{i}
-\bar{c}^{a}c^{a}\right)\rangle$ \cite{prep}, $v^2 \sim \langle
\e^{ab} \bar c^a c^b \rangle$ \cite{Capri:2007hw}.
\\\\The action \eqref{clact} constitutes our starting point in
order to establish the renormalizability of the model.
\section{Ward identities}
\label{ward_ids}
It turns out that $\S$ fulfills the following set of Ward
identities:
\begin{itemize}
\item{The Slavnov-Taylor identity:
\begin{eqnarray}
\mathcal{S}(\S)&\equiv&\Int\left[\left(\frac{\d\S}{\d\Omega^{a}_{\mu}}
+\frac{\d\S}{\d K^{a}_{\mu}}\right)\frac{\d\S}{\d A^{a}_{\mu}}
+\frac{\d\S}{\d\Omega_{\mu}}\frac{\d\S}{\d A_{\mu}}
+\frac{\d\S}{\d L^{a}}\frac{\d\S}{\d c^{a}} +\frac{\d\S}{\d
L}\frac{\d\S}{\d c} +ib^{a}\frac{\d\S}{\d\bar{c}^{a}}
+ib\frac{\d\S}{\d\bar{c}}\right.\nonumber\\
&&+\omega^{a}_{i}\frac{\d\S}{\d\phi^{a}_{i}}
+\bar{\phi}^{a}_{i}\frac{\d\S}{\d\bar{\omega}^{a}_{i}} +N^{a}_{\mu
i}\frac{\d\S}{\d M^{a}_{\mu i}}-\bar{M}^{a}_{\mu
i}\frac{\d\S}{\d\bar{N}^{a}_{\mu i}}
-(\Omega^{a}_{\mu}-K^{a}_{\mu})\frac{\d\S}{\d\xi^{a}_{\mu}}
-\bar{Y}^{a}_{i}\frac{\d\S}{\d\bar{X}^{a}_{i}}\nonumber\\
&&\left.+X^{a}_{i}\frac{\d\S}{\d Y^{a}_{i}} +J\frac{\d\S}{\d\lambda}
+\s\frac{\d\S}{\d\tau}+\theta\frac{\d\S}{\d \eta}
-\theta_{i}\frac{\d\S}{\d \eta_{i}}
+\bar{\eta}_{i}\frac{\d\S}{\d\bar{\theta}_{i}}\,\right]=0\;.
\end{eqnarray}}

\item{The four global $\mathcal{W}^{(N)}_{i}$-identities which mix
the Faddev-Popov ghost fields with the auxiliary localizing
fields:
\begin{equation}
\mathcal{W}^{(N)}_{i}(\S)=0\;,\qquad (N=1,2,3,4)\;,
\end{equation}
where
\begin{eqnarray}
\mathcal{W}^{(1)}_{i}\!(\S)&\!\!\!\!\equiv\!\!\!\!&\Int\!\!\left(
\phi^{a}_{i}\frac{\d\S}{\d\bar{c}^{a}}
+c^{a}\frac{\d\S}{\d\bar{\phi}^{a}_{i}} +M^{a}_{\mu
i}\frac{\d\S}{\d\Omega^{a}_{\mu}} -Y^{a}_{i}\frac{\d\S}{\d L^{a}}
-i\frac{\d\S}{\d\bar{Y}^{a}_{i}}\frac{\d\S}{\d b^{a}}
-\theta\frac{\d\S}{\d\bar{\eta}_{i}} +2\eta_{i}\frac{\d\S}{\d L}\right)\,,\nonumber\\
\mathcal{W}^{(2)}_{i}\!(\S)&\!\!\!\!\equiv\!\!\!\!&\Int\!\!\left(
\bar{\omega}^{a}_{i}\frac{\d\S}{\d\bar{c}^{a}}
-c^{a}\frac{\d\S}{\d\omega^{a}_{i}} -\bar{N}^{a}_{\mu
i}\frac{\d\S}{\d\Omega^{a}_{\mu}} -\bar{X}^{a}_{i}\frac{\d\S}{\d
L^{a}} -i\frac{\d\S}{\d X^{a}_{i}}\frac{\d\S}{\d b^{a}}
-\theta\frac{\d\S}{\d \theta_{i}} +2\bar{\theta}_{i}\frac{\d\S}{\d L}\right)\,,\nonumber\\
\mathcal{W}^{(3)}_{i}\!(\S)&\!\!\!\!\equiv\!\!\!\!&\Int\!\!\left[
\left(\frac{\d\S}{\d\bar{Y}^{a}_{i}}
+\omega^{a}_{i}\right)\frac{\d\S}{\d\bar{c}^{a}}
+i\frac{\d\S}{\d\bar{X}^{a}_{i}}\frac{\d\S}{\d b^{a}}
+\left(\frac{\d\S}{\d\bar{\phi}^{a}_{i}}-X^{a}_{i}\right)
\frac{\d\S}{\d L^{a}} +c^{a}\frac{\d\S}{\d\bar{\omega}^{a}_{i}}\right.\nonumber\\
&&\left. -M^{a}_{\mu i}\frac{\d\S}{\d\xi^{a}_{\mu}} +N^{a}_{\mu
i}\frac{\d\S}{\d\Omega^{a}_{\mu}}
-\theta\frac{\d\S}{\d\bar{\theta}_{i}} -2\theta_{i}\frac{\d\S}{\d L}\,\right]\,,\nonumber\\
\mathcal{W}^{(4)}_{i}\!(\S)&\!\!\!\!\equiv\!\!\!\!&\Int\!\!\left[
\left(\frac{\d\S}{\d X^{a}_{i}}-\bar{\phi}^{a}_{i}\right)
\frac{\d\S}{\d\bar{c}^{a}} -i\frac{\d\S}{\d
Y^{a}_{i}}\frac{\d\S}{\d b^{a}}
+\left(\frac{\d\S}{\d\omega^{a}_{i}}-\bar{Y}^{a}_{i}\right)
\frac{\d\S}{\d L^{a}} -c^{a}\frac{\d\S}{\d\phi^{a}_{i}}\right.\nonumber\\
&&\left. +\bar{N}^{a}_{\mu i}\frac{\d\S}{\d\xi^{a}_{\mu}}
-\bar{M}^{a}_{\mu i}\frac{\d\S}{\d\Omega^{a}_{\mu}}
+\theta\frac{\d\S}{\d \eta_{i}} -2\bar{\eta}_{i}\frac{\d\S}{\d
L}\,\right]\,.\label{W_symm}
\end{eqnarray}}
\item{The global $U(8)$ invariance:
\begin{equation}
{\cal Q}_{ij}(\S)=0\;,
\end{equation}
with
\begin{eqnarray}
{\cal Q}_{ij}&\!\!\!\!\!\!=\!\!\!\!\!\!&\int
d^{4}\!x\,\biggl(\phi^{a}_{i}\frac{\delta}{\delta\phi^{a}_{j}}
-\bar\phi^{a}_{j}\frac{\delta}{\delta\bar\phi^{a}_{i}}+
\omega^{a}_{i}\frac{\delta}{\delta\omega^{a}_{j}}
-\bar\omega^{a}_{j}\frac{\delta}{\delta\bar\omega^{a}_{i}}
+M^{a}_{\mu i}\frac{\delta}{\delta M^{a}_{\mu j}} -\bar{M}^{a}_{\mu
j}\frac{\delta}{\delta \bar{M}^{a}_{\mu i}} +N^{a}_{\mu
i}\frac{\delta}{\delta N^{a}_{\mu j}}
\nonumber\\
&&-\bar{N}^{a}_{\mu j}\frac{\delta}{\delta \bar{N}^{a}_{\mu i}}
+Y^{a}_{i}\frac{\d}{\d Y^{a}_{j}}
-\bar{Y}^{a}_{j}\frac{\d}{\d\bar{Y}^{a}_{i}} +X^{a}_{i}\frac{\d}{\d
X^{a}_{j}} -\bar{X}^{a}_{j}\frac{\d}{\d\bar{X}^{a}_{i}}
+\eta_{i}\frac{\d}{\d \eta_{j}}
-\bar{\eta}_{j}\frac{\d}{\d\bar{\eta}_{i}} +\theta_{i}\frac{\d}{\d
\theta_{j}}
-\bar{\theta}_{j}\frac{\d}{\d\bar{\theta}_{i}}\biggr)\;.\label{Q_8}
\end{eqnarray}
The trace of \eqref{Q_8} defines a new charge displayed in the
Tables \ref{table1} and \ref{table2}. This operator generalizes
that of eq.\eqref{q8}}.

\begin{table}[t]\centering
{\small\begin{tabular}{lcccccccccccccccccccccc} \hline\hline
&$\Omega$&$K$&$\xi$&$L$&$Y$&$\bar{Y}$&$X$&$\bar{X}$&$\lambda$&$\tau$&$J$&$\s$&$\eta$&$\theta$
&$\eta_{i}$&$\bar{\eta}_{i}$&$\theta_{i}$&
$\bar{\theta}_{i}\!\phantom{\Bigl|}$\\
\hline
dim&$3$&$3$&$3$&$4$&$3$&$3$&$3$&$3$&$2$&$2$&$2$&$2$&$2$&$2$&$3$&$3$&$3$&$3$\\
gh. number&$-1$&$-1$&$-2$&$-2$&$-1$&$-1$&$0$&$-2$&$-1$&$-1$&$0$&$0$&$-1$&$0$&$-1$&$-1$&$0$&$-2$\\
$\mathcal{Q}_{8}$-charge&$0$&$0$&$0$&$0$&$1$&$-1$&$1$&$-1$&$0$&$0$&$0$&$0$&$0$&$0$&$1$&$-1$&$1$&$-1$\\
\hline\hline
\end{tabular}}
\caption{Quantum numbers of the external sources} \label{table2}
\end{table}

\item{The exact rigid symmetries:
\begin{eqnarray}
\mathfrak{R}_{ij}(\S)&\equiv&\int
d^{4}\!x\,\biggr(\phi^{a}_{i}\frac{\delta \S}{\delta\omega^{a}_{j}}
-\bar\omega^{a}_{j}\frac{\delta
\S}{\delta\bar\phi^{a}_{i}}+M^{a}_{\mu i}\frac{\delta \S}{\delta
N^{a}_{\mu j}}+\bar{N}^{a}_{\mu j}\frac{\delta \S}{\delta
\bar{M}^{a}_{\mu i}} +Y^{a}_{i}\frac{\delta \S}{\delta X^{a}_{j}}
+\bar{X}^{a}_{j}\frac{\delta \S}{\delta\bar{Y}^{a}_{i}}\nonumber\\
&& -\eta_{i}\frac{\d\S}{\d\theta_{j}}
-\bar{\theta}_{j}\frac{\d\S}{\d\bar{\eta}_{i}}\biggl)=0\;,\nonumber\\
\mathfrak{R}^{(1)}(\S)&\equiv&\int
d^{4}\!x\,\biggl(\bar\omega^{a}_{i} \frac{\delta
\S}{\delta\omega^{a}_{i}}-\bar{N}^{a}_{\mu i} \frac{\delta
\S}{\delta N^{a}_{\mu i}}+\bar{X}^{a}_{i}\frac{\delta
\S}{\delta X^{a}_{i}} -\bar{\theta}_{i}\frac{\d\S}{\d \theta_{i}}\biggr)=0\;,\nonumber\\
\mathfrak{R}^{(2)}(\S)&\equiv&\int d^{4}\!x\,\biggl(
\bar\omega^{a}_{i}\frac{\delta{\S}}{\delta\phi^{a}_{i}}
-\bar\phi^{a}_{i}\frac{\delta{\S}}{\delta\omega^{a}_{i}} -
\bar{N}^{a}_{\mu i}\frac{\delta{\S}}{\delta M^{a}_{\mu
i}}-\bar{M}^{a}_{\mu i}\frac{\delta{\S}}{\delta N^{a}_{\mu
i}}+\bar{X}^{a}_{i}\frac{\delta{\S}}{\delta Y^{a}_{i}}
+\bar{Y}^{a}_{i}\frac{\delta{\S}}{\delta
X^{a}_{i}}\nonumber\\
&&+\bar{\theta}_{i}\frac{\d\S}{\d\eta_{i}}
+\bar{\eta}_{i}\frac{\d\S}{\d\theta_{i}}\biggr)=0\;.
\end{eqnarray}}

\item{The diagonal gauge fixing condition:
\begin{equation}
\frac{\d\S}{\d b} = i\p_{\mu}A_{\mu}\;.
\end{equation}}

\item{The diagonal anti-ghost equation:
\begin{equation}
\frac{\d\S}{\d\bar{c}}+\p_{\mu}\frac{\d\S}{\d\Omega_{\mu}}=0\;.
\end{equation}}

\item{The $SL(2,\mathbb{R})$ symmetry:
\begin{equation}
\mathcal{D}(\S)\equiv\Int\left( c^{a}\frac{\d\S}{\d\bar{c}^{a}}
-i\frac{\d\S}{\d L^{a}}\frac{\d\S}{\d b^{a}}
-2\theta\frac{\d\S}{\d L}\right)=0\;.\label{Dsymm}
\end{equation}}

\item{The local $U(1)$ invariance:
\begin{equation}
\mathcal{W}^{3}(\S)=-i\p^{2}b\;,
\end{equation}
with
\begin{equation}
\mathcal{W}^{3}\equiv\p_{\mu}\frac{\d}{\d A_{\mu}}
+g\e^{ab}\sum_{\mathcal{Y}\in\mathrm{\textsl{O\hspace{-0.5pt}f\hspace{0.02pt}f}}
}\mathcal{Y}^{a}\frac{\d}{\d \mathcal{Y}^{b}},
\end{equation}
where
\begin{equation}
\mathrm{\textsl{O\hspace{-0.6pt}f\hspace{0.02pt}f}}=
\Bigl\{A^{a}_{\mu},b^{a},\bar{c}^{a},c^{a},
\bar{\phi}^{a}_{i},\phi^{a}_{i},\bar{\omega}^{a}_{i},\omega^{a}_{i},
\Omega^{a}_{\mu},K^{a}_{\mu},\xi^{a}_{\mu}, L^{a},
\bar{X}^{a}_{i},X^{a}_{i},\bar{Y}^{a}_{i},Y^{a}_{i},
\bar{M}^{a}_{\mu i},M^{a}_{\mu i},\bar{N}^{a}_{\mu i},N^{a}_{\mu
i}\Bigr\}\;.\label{off_diagonal_set}
\end{equation}}

\item{The BRST on-shell invariance of the general operator
$(\mathcal{O}_{A^{2}} +\alpha\,\mathcal{O}_{\bar{f}f})$:
\begin{equation}
\mathcal{U}(\S)\equiv\Int \left(\frac{\d\S}{\d\lambda}
+\alpha\frac{\d\S}{\d\tau} -ic^{a}\frac{\d\S}{\d b^{a}}
-2\eta\frac{\d\S}{\d L}\right)=0\;.
\end{equation}}

\end{itemize}

\section{Algebraic characterization of the most general counterterm}
We can face now the issue of the renormalizability of the starting
action $\S$. We shall employ the algebraic renormalization
\cite{Piguet:1995er} and look for the most general invariant
counterterm which can be freely added to all orders of
perturbation theory. To that purpose we perturb the classical
action $\S$ by adding an arbitrary integrated local polynomial
$\S_{\mathrm{CT}}$ in the fields and external sources of dimension
bounded by four, zero ghost number and zero
$\mathcal{Q}_{8}$-charge. Requiring thus the perturbed action,
$\S+\epsilon\S_{\mathrm{CT}}$, satisfies the same Ward identities
as $\S$ to the first order in the perturbation parameter
$\epsilon$, we get:
\begin{eqnarray}
\mathcal{S}(\S+\epsilon\S_{\mathrm{CT}})&=&0+O(\epsilon^{2})\;,\nonumber\\
\frac{\d}{\d b}(\S+\epsilon\S_{\mathrm{CT}})&=&i\p_{\mu}A_{\mu}+O(\epsilon^{2})\;,\nonumber\\
\Bigl(\,\frac{\d}{\d\bar{c}}
+\p_{\mu}\frac{\d}{\d\Omega_{\mu}}\,\Bigr)
(\S+\epsilon\S_{\mathrm{CT}})&=&0+O(\epsilon^{2})\;,\nonumber\\
\mathcal{W}^{(N)}_{i}(\S+\epsilon\S_{\mathrm{CT}})&=&
0+O(\epsilon^{2})\;,\qquad(N=1,2,3,4)\;,\nonumber\\
\mathfrak{R}_{ij}(\S+\epsilon\S_{\mathrm{CT}})
&=&0+O(\epsilon^{2})\;,\nonumber\\
\mathfrak{R}^{(K)}(\S+\epsilon\S_{\mathrm{CT}})
&=&0+O(\epsilon^{2})\;,\qquad(K=1,2)\;,\nonumber\\
\mathcal{Q}_{ij}(\S+\epsilon\S_{\mathrm{CT}})
&=&0+O(\epsilon^{2})\;,\nonumber\\
\mathcal{D}(\S+\epsilon\S_{\mathrm{CT}})
&=&0+O(\epsilon^{2})\;,\nonumber\\
\mathcal{W}^{3}(\S+\epsilon\S_{\mathrm{CT}})
&=&-i\p^{2}b+O(\epsilon^{2})\;,\nonumber\\
\mathcal{U}(\S+\epsilon\S_{\mathrm{CT}}) &=&0+O(\epsilon^{2})\;.
\end{eqnarray}
This amounts to imposing the following conditions on
$\S_{\mathrm{CT}}$
\begin{eqnarray}
\mathcal{S}_{\S}\S_{\mathrm{CT}}&=&0\;,\nonumber\\
\frac{\d}{\d b}\S_{\mathrm{CT}}&=&0\;,\nonumber\\
\Bigl(\,\frac{\d}{\d\bar{c}}
+\p_{\mu}\frac{\d}{\d\Omega_{\mu}}\,\Bigr)
\S_{\mathrm{CT}}&=&0\;,\nonumber\\
\mathcal{W}^{\S(N)}_{i}\S_{\mathrm{CT}}&=&
0\;,\qquad(N=1,2,3,4)\;,\nonumber\\
\mathfrak{R}_{ij}\S_{\mathrm{CT}}
&=&0\;,\nonumber\\
\mathfrak{R}^{(K)}\S_{\mathrm{CT}}
&=&0\;,\qquad(K=1,2)\;,\nonumber\\
\mathcal{Q}_{ij}\S_{\mathrm{CT}}
&=&0\;,\nonumber\\
\mathcal{D}_{\S}\S_{\mathrm{CT}}
&=&0\;,\nonumber\\
\mathcal{W}^{3}\S_{\mathrm{CT}}
&=&0\;,\nonumber\\
\mathcal{U}\,\S_{\mathrm{CT}} &=&0\;,\label{constraints}
\end{eqnarray}
where $\mathcal{S}_{\S}$ is the nilpotent linearized
Slavnov-Taylor operator,
\begin{equation}
\mathcal{S}_{\S}\mathcal{S}_{\S}=0\;,
\end{equation}
\begin{eqnarray}
\mathcal{S}_{\S}&=&\Int\left[\left(\frac{\d\S}{\d\Omega^{a}_{\mu}}
+\frac{\d\S}{\d K^{a}_{\mu}}\right)\frac{\d}{\d A^{a}_{\mu}}
+\frac{\d\S}{\d A^{a}_{\mu}}\left(\frac{\d}{\d\Omega^{a}_{\mu}}
+\frac{\d}{\d K^{a}_{\mu}}\right)
+\frac{\d\S}{\d\Omega_{\mu}}\frac{\d}{\d A_{\mu}} +\frac{\d\S}{\d
A_{\mu}}\frac{\d}{\d\Omega_{\mu}}
+\frac{\d\S}{\d L^{a}}\frac{\d}{\d c^{a}}\right.\nonumber\\
&&+\frac{\d\S}{\d c^{a}}\frac{\d}{\d L^{a}} +\frac{\d\S}{\d
L}\frac{\d}{\d c} +\frac{\d\S}{\d c}\frac{\d}{\d L}
+ib^{a}\frac{\d}{\d\bar{c}^{a}} +ib\frac{\d}{\d\bar{c}}
+\omega^{a}_{i}\frac{\d}{\d\phi^{a}_{i}}
+\bar{\phi}^{a}_{i}\frac{\d}{\d\bar{\omega}^{a}_{i}} +N^{a}_{\mu
i}\frac{\d}{\d M^{a}_{\mu i}} -\bar{M}^{a}_{\mu
i}\frac{\d}{\d\bar{N}^{a}_{\mu i}}\nonumber\\
&&\left.-(\Omega^{a}_{\mu}-K^{a}_{\mu})\frac{\d}{\d\xi^{a}_{\mu}}
-\bar{Y}^{a}_{i}\frac{\d}{\d\bar{X}^{a}_{i}}
+X^{a}_{i}\frac{\d}{\d Y^{a}_{i}} +J\frac{\d}{\d\lambda}
+\s\frac{\d}{\d\tau} +\theta\frac{\d}{\d\eta}
-\theta_{i}\frac{\d}{\d\eta_{i}}
+\bar{\eta}_{i}\frac{\d}{\d\bar{\theta}_{i}}\,\right]\;,
\end{eqnarray}
while $\mathcal{W}^{\S(N)}_{i}$, with $N=1,\dots,4$, and
$\mathcal{D}_{\S}$ are the linearized operators corresponding to
the Ward identities \eqref{W_symm} and \eqref{Dsymm},
respectively, and they are given by
\begin{eqnarray}
\mathcal{W}^{\S(1)}_{i}&=&\Int\left(
\phi^{a}_{i}\frac{\d}{\d\bar{c}^{a}}
+c^{a}\frac{\d}{\d\bar{\phi}^{a}_{i}} +M^{a}_{\mu
i}\frac{\d}{\d\Omega^{a}_{\mu}} -Y^{a}_{i}\frac{\d}{\d L^{a}}
-i\frac{\d\S}{\d\bar{Y}^{a}_{i}}\frac{\d}{\d b^{a}}
-i\frac{\d\S}{\d b^{a}}\frac{\d}{\d\bar{Y}^{a}_{i}}
-\theta\frac{\d}{\d\bar{\eta}_{i}} +2\eta_{i}\frac{\d}{\d
L}\right)\;,\nonumber\cr \mathcal{W}^{\S(2)}_{i}&=&\Int\left(
\bar{\omega}^{a}_{i}\frac{\d}{\d\bar{c}^{a}}
-c^{a}\frac{\d}{\d\omega^{a}_{i}} -\bar{N}^{a}_{\mu
i}\frac{\d}{\d\Omega^{a}_{\mu}} -\bar{X}^{a}_{i}\frac{\d}{\d
L^{a}} -i\frac{\d\S}{\d X^{a}_{i}}\frac{\d}{\d b^{a}}
-i\frac{\d\S}{\d b^{a}}\frac{\d}{\d X^{a}_{i}}
-\theta\frac{\d}{\d\theta_{i}} +2\bar{\theta}_{i}\frac{\d}{\d
L}\right)\;,\nonumber\cr \mathcal{W}^{\S(3)}_{i}&=&\Int\left[
\left(\frac{\d\S}{\d\bar{Y}^{a}_{i}}
+\omega^{a}_{i}\right)\frac{\d}{\d\bar{c}^{a}}
+\frac{\d\S}{\d\bar{c}^{a}}\frac{\d}{\d\bar{Y}^{a}_{i}}
+i\frac{\d\S}{\d\bar{X}^{a}_{i}}\frac{\d}{\d b^{a}}
+i\frac{\d\S}{\d b^{a}}\frac{\d}{\d\bar{X}^{a}_{i}}
+\left(\frac{\d\S}{\d\bar{\phi}^{a}_{i}}-X^{a}_{i}\right)
\frac{\d}{\d L^{a}}\right.\nonumber\\
&&\left.+\frac{\d\S}{\d L^{a}}\frac{\d}{\d\bar{\phi}^{a}_{i}}
+c^{a}\frac{\d}{\d\bar{\omega}^{a}_{i}} -M^{a}_{\mu
i}\frac{\d}{\d\xi^{a}_{\mu}} +N^{a}_{\mu
i}\frac{\d}{\d\Omega^{a}_{\mu}}
-\theta\frac{\d}{\d\bar{\theta}_{i}} -2\theta_{i}\frac{\d}{\d
L}\,\right]\;,\nonumber\cr \mathcal{W}^{\S(4)}_{i}&=&\Int\left[
\left(\frac{\d\S}{\d X^{a}_{i}}-\bar{\phi}^{a}_{i}\right)
\frac{\d}{\d\bar{c}^{a}} +\frac{\d\S}{\d\bar{c}^{a}}\frac{\d}{\d
X^{a}_{i}} -i\frac{\d\S}{\d Y^{a}_{i}}\frac{\d}{\d b^{a}}
-i\frac{\d\S}{\d b^{a}}\frac{\d}{\d Y^{a}_{i}}
+\left(\frac{\d\S}{\d\omega^{a}_{i}}-\bar{Y}^{a}_{i}\right)
\frac{\d}{\d L^{a}}\right.\nonumber\\
&&\left. +\frac{\d\S}{\d L^{a}}\frac{\d}{\d\omega^{a}_{i}}
-c^{a}\frac{\d}{\d\phi^{a}_{i}} +\bar{N}^{a}_{\mu
i}\frac{\d}{\d\xi^{a}_{\mu}} -\bar{M}^{a}_{\mu
i}\frac{\d}{\d\Omega^{a}_{\mu}} +\theta\frac{\d}{\d\eta_{i}}
-2\bar{\eta}_{i}\frac{\d}{\d L}\,\right]\;,
\end{eqnarray}
and
\begin{equation}
\mathcal{D}_{\S}=\Int\left( c^{a}\frac{\d}{\d\bar{c}^{a}}
-i\frac{\d\S}{\d L^{a}}\frac{\d}{\d b^{a}} -i\frac{\d\S}{\d
b^{a}}\frac{\d}{\d L^{a}} -2\theta\frac{\d}{\d L}\right)\;.
\end{equation}
For further use, let us write some useful commutation and
anticommutation relations
\begin{equation}
\begin{tabular}{rclrcl}
$\left\{\mathcal{W}^{\S(1)}_{i},\mathcal{S}_{\S}\right\}$&$\!\!\!=\!\!\!$&$
\mathcal{W}^{\S(3)}_{i}\,,\qquad$
&$\left[\mathcal{W}^{\S(2)}_{i},\mathcal{S}_{\S}\right]$
&$\!\!\!=\!\!\!$&$
\mathcal{W}^{\S(4)}_{i}\,,$\vspace{3pt}\\

$\left\{\mathfrak{R}_{ij},\mathcal{S}_{\S}\right\}$&$\!\!\!=\!\!\!$
&$\mathcal{Q}_{ij}\,,$
&$\left[\mathfrak{R}^{(1)},\mathcal{S}_{\S}\right]$&$\!\!\!=\!\!\!$
&$\mathfrak{R}^{(2)}\,,\vspace{3pt}$\\

$\left[\mathcal{W}^{3},\mathcal{S}_{\S}\right]$&$\!\!\!=\!\!\!$&$0\,,$
&$\left\{\mathcal{U},\mathcal{S}_{\S}\right\}$&$\!\!\!=\!\!\!$
&$\mathcal{D}_{\S}\,.\vspace{3pt}$\\

\end{tabular}
\end{equation}
From the second and the third constraints of \eqref{constraints}
it follows that $\S_{\mathrm{CT}}$ is independent from the
diagonal Lagrange multiplier $b$, and that the diagonal antighost
$\bar{c}$ enters only through the combination $(\Omega_{\mu} +
\p_{\mu}\bar{c})$. Furthermore, from general results on the
cohomology of gauge theories \cite{Piguet:1995er}, it turns out
that the most general solution of the constraint
$\mathcal{S}_{\S}\S_{\mathrm{CT}}=0$, {\it i.e.} the first of
eqs.\eqref{constraints}, can be written as
\begin{equation}
\S_{\mathrm{CT}}=a_{0}\,S_{\mathrm{YM}}
+\mathcal{S}_{\S}\Delta^{(-1)}\;,
\end{equation}
with $\Delta^{(-1)}$ being an integrated local polynomial with
ghost number $-1$, given by
\begin{eqnarray}
\Delta^{(-1)}&=&\Int\Bigl[a_{1}\,\Omega^{a}_{\mu}A^{a}_{\mu}+
a_{2}\,K^{a}_{\mu}A^{a}_{\mu}+
a_{3}\,\xi^{a}_{\mu}\,g\varepsilon^{ab\!}A^{b}_{\mu}c+
a_{4}\,\xi^{a}_{\mu}\,\partial_{\mu}c^{a}+
a_{5}\,\xi^{a}_{\mu}\,g\varepsilon^{ab\!}A_{\mu}c^{b}+
a_{6}\,(\partial_{\mu}\bar c^{a})A^{a}_{\mu}\nonumber\\
&&+a_{7}\,(\Omega_{\mu}+\partial_{\mu}\bar
c)A_{\mu}+a_{8}\,c^{a}L^{a}+a_{9}\,cL+
a_{10}\,\bar{X}^{a}_{i}\,g\varepsilon^{ab\!}\phi^{b}_{i}c+
a_{11}\,\bar{X}^{a}_{i}\omega^{a}_{i}+
a_{12}\,Y^{a}_{i}\,g\varepsilon^{ab\!}\bar\omega^{b}_{i}c
\phantom{\Bigr|}\nonumber\\
&&+a_{13}\,Y^{a}_{i}\bar\phi^{a}_{i}+
a_{14}\,\bar{Y}^{a}_{i}\phi^{a}_{i}+a_{15}\,X^{a}_{i}\bar\omega^{a}_{i}+
a_{16}\,\bar{N}^{a}_{\mu i}\,\partial_{\mu}\phi^{a}_{i}+
a_{17}\,\bar{N}^{a}_{\mu
i}\,g\varepsilon^{ab\!}A_{\mu}\phi^{b}_{i}+ a_{18}M^{a}_{\mu
i}\,\partial_{\mu}\bar\omega^{a}_{i}
\phantom{\Bigr|}\nonumber\\
&&+a_{19}\,M^{a}_{\mu
i}\,g\varepsilon^{ab\!}A_{\mu}\bar\omega^{b}_{i} +a_{20}\,
ib^{a}\bar c^{a} +a_{21}\,g\varepsilon^{ab\!}\bar c^{a}\bar c^{b}
c +a_{22}\,g\varepsilon^{ab\!}\bar c^{a}A_{\mu}A^{b}_{\mu}
+a_{23}\,\bar\omega^{a}_{i}\phi^{a}_{i}\bar\phi^{b}_{j}\phi^{b}_{j}
\phantom{\Bigr|}\nonumber\\
&&+a_{24}\,\bar\omega^{a}_{i}\phi^{a}_{i}\bar\omega^{b}_{j}\omega^{b}_{j}
+a_{25}\,\bar\omega^{a}_{i}\phi^{b}_{i}\bar\phi^{a}_{j}\phi^{b}_{j}
+a_{26}\,\bar\omega^{a}_{i}\phi^{b}_{i}\bar\omega^{a}_{j}\omega^{b}_{j}
+a_{27}\,\bar\omega^{a}_{i}\phi^{b}_{i}\bar\phi^{b}_{j}\phi^{a}_{j}
+a_{28}\,\bar\omega^{a}_{i}\phi^{b}_{i}\bar\omega^{b}_{j}\omega^{a}_{j}
\phantom{\Bigl|}\nonumber\\
&&+a_{29}\,\bar\omega^{a}_{i}\phi^{a}_{j}\bar\phi^{b}_{i}\phi^{b}_{j}
+a_{30}\,\bar\omega^{a}_{i}\phi^{b}_{j}\bar\phi^{a}_{i}\phi^{b}_{j}
+a_{31}\bar\omega^{a}_{i}\phi^{a}_{j}\bar\omega^{b}_{i}\omega^{b}_{j}
+a_{32}\,\bar\omega^{a}_{i}\phi^{a}_{i}\bar c^{b}c^{b}
+a_{33}\,\bar\omega^{a}_{i}\phi^{b}_{i}\bar c^{a}c^{b}
\phantom{\Bigl|}\nonumber\\
&&+a_{34}\,\bar\omega^{a}_{i}\phi^{b}_{i}\bar c^{b}c^{a}
+a_{35}\,\bar\omega^{a}_{i}\phi^{a}_{i}A_{\mu}A_{\mu}
+a_{36}\,\bar\omega^{a}_{i}\phi^{a}_{i}A^{b}_{\mu}A^{b}_{\mu}
+a_{37}\,\bar\omega^{a}_{i}\phi^{b}_{i}A^{a}_{\mu}A^{b}_{\mu}
+a_{38}\,\bar\omega^{a}_{i}\partial^{2}\phi^{a}_{i}
\phantom{\Bigl|}\nonumber\\
&&+a_{39}\,\bar\omega^{a}_{i}\,g\varepsilon^{ab\!}
A_{\mu}\partial_{\mu}\phi^{b}_{i}
+a_{40}\,\bar\omega^{a}_{i}\,g\varepsilon^{ab\!}
(\partial_{\mu}A_{\mu})\phi^{b}_{i} +a_{41}\,\chi \bar{N}^{a}_{\mu
i}M^{a}_{\mu i}
+\half(a_{42}\,\lambda+a_{43}\,\tau)A^{a}_{\mu}A^{a}_{\mu}
\phantom{\Bigl|}\nonumber\\
&&+\half(a_{44}\,\lambda+a_{45}\,\tau)A_{\mu}A_{\mu}
+(a_{46}\,\lambda+a_{47}\,\tau)\bar{c}^{a}c^{a}
+(a_{48}\,\lambda+a_{49}\,\tau)\bar{\phi}^{a}_{i}\phi^{a}_{i}
+(a_{50}\,\lambda+a_{51}\,\tau)\bar{\omega}^{a}_{i}\omega^{a}_{i}
\phantom{\Bigl|}\nonumber\\
&&+(a_{52}\,J+a_{53}\,\s)\bar{\omega}^{a}_{i}\phi^{a}_{i}
+(a_{54}\,J+a_{55}\,\s)\lambda +(a_{56}\,J+a_{57}\,\s)\tau
+a_{58}\,\eta\,\p_{\mu}A_{\mu}
+a_{59}\,\eta\,\theta \nonumber\\
&&+g\e^{ab}\left(a_{60}\,\eta\,\bar{c}^{a}c^{b}
+a_{61}\,\eta\,\bar{\phi}^{a}_{i}\phi^{b}_{i}
+a_{62}\,\eta\,\bar{\omega}^{a}_{i}\omega^{b}_{i}
+a_{63}\,\theta\,\bar{\omega}^{a}_{i}\phi^{b}_{i}
+a_{64}\,\bar{\theta}_{i}\phi^{a}_{i}c^{b}
+a_{65}\,\eta_{i}\bar{\omega}^{a}_{i}c^{b}\right)\Bigr]\;,\label{delta-1}
\end{eqnarray}
where the coefficients $a_{n}$, $n=0,\dots,65$, are free
dimensionless parameters. Notice also that in the derivation of
expression \eqref{delta-1} use has been made of the fact that the
action $\S$, and thus $\S_{\mathrm{CT}}$, are left invariant by
the following discrete symmetry
\begin{equation}
\mathcal{Y}^{1}\to \mathcal{Y}^{1},\qquad \mathcal{Y}^{2}\to
-\mathcal{Y}^{2},\qquad \mathcal{Y}^{\mathrm{\,diag}}\to
-\mathcal{Y}^{\mathrm{\,diag}},
\qquad\mathcal{Y}\to\mathcal{Y},\label{discrete}
\end{equation}
where $\mathcal{Y}^{a}$, with $a=1,2$, stands for the elements of
the off-diagonal set \eqref{off_diagonal_set}, while
$\mathcal{Y}^{\mathrm{\,diag}}$ for the diagonal sector
\begin{equation}
\mathcal{Y}^{\mathrm{\,diag}}\in\Bigl\{A_{\mu},b,c,\bar{c},
\Omega_{\mu},L,\eta,\theta,\eta_{i},\bar{\eta}_{i},\theta_{i},\bar{\theta}_{i}\Bigr\}\;,
\end{equation}
and $\mathcal{Y}$ the sources $\lambda$, $\tau$, $J$, $\s$. As one
can easily recognize, this symmetry plays the role of the charge
conjugation. \\\\After a quite lengthy calculation one finds that
the most general expression for $\Delta^{(-1)}$ compatible with
all constraints \eqref{constraints} and the discrete symmetry
\eqref{discrete} is
\begin{eqnarray}
\Delta^{(-1)}&\!\!\!=\!\!\!&\Int\Bigl\{(a_{1}+a_{2})\bigl(\Omega^{a}_{\mu}A^{a}_{\mu}
+g\e^{ab}\xi^{a}_{\mu}A^{b}_{\mu}c\bigr)
+(a_{2}+a_{4})\xi^{a}_{\mu}D^{ab}_{\mu}c^{b}
+(a_{1}-a_{4}+a_{8})\bigl(\bar{N}^{a}_{\mu
i}D^{ab}_{\mu}\phi^{b}_{i}
-M^{a}_{\mu i}D^{ab}_{\mu}\bar{\omega}^{b}_{i}\bigr)\nonumber\\
&&-(a_{6}+a_{8})\bar{\omega}^{a}_{i}\mathcal{M}^{ab}\phi^{b}_{i}
-a_{6}\,\bar{c}^{a}D^{ab}_{\mu}A^{b}_{\mu} +a_{8}\,L^{a}c^{a}
+a_{9}\,Lc +a_{41}\,\chi\,\bar{N}^{a}_{\mu i}M^{a}_{\mu i}
+a_{20}(i\bar{c}^{a}b^{a} -g\e^{ab}\bar{c}^{a}\bar{c}^{b}c)\nonumber\\
&&-(-\alpha a_{8}-a_{20})g^{2}\bar{\omega}^{a}_{i}\phi^{a}_{i}
(\bar{\phi}^{b}_{j}\phi^{b}_{j}
-\bar{\omega}^{b}_{j}\omega^{b}_{j}) +(-\alpha
a_{8}-2a_{20})g^{2}\bar{\omega}^{a}_{i}\phi^{a}_{i}\bar{c}^{b}c^{b}
+\half(a_{8}-a_{6}-\alpha a_{43})\lambda\, A^{a}_{\mu}A^{a}_{\mu}\nonumber\\
&&+\half a_{43}\,\tau A^{a}_{\mu}A^{a}_{\mu} -(-\alpha
a_{8}+2a_{20}-\alpha a_{47})\,\lambda\bigl(
\bar{\phi}^{a}_{i}\phi^{a}_{i} -\bar{\omega}^{a}_{i}\omega^{a}_{i}
-\bar{c}^{a}c^{a}\bigr)-(a_{8}+a_{47})\,\tau \bigl(
\bar{\phi}^{a}_{i}\phi^{a}_{i}
-\bar{\omega}^{a}_{i}\omega^{a}_{i}\bigr)
+a_{47}\,\tau\bar{c}^{a}c^{a}
\nonumber\\
&&-(a_{8}-a_{9})g\e^{ab}(\eta\bar{c}^{a}c^{b}
+\bar{\theta}_{i}\phi^{a}_{i}c^{b}
-\eta_{i}\bar{\omega}^{a}c^{b})+a_{54}\,J\lambda
+a_{55}\,\s\lambda +a_{56}\,J\tau +a_{57}\,\s\tau
+a_{59}\,\eta\theta\Bigr\}\;.
\end{eqnarray}
Noticing that
\begin{equation}
J\tau =\s\lambda+\mathcal{S}_{\S}(\lambda\tau)\;,
\end{equation}
and renaming the coefficients as
\begin{eqnarray}
&a_{1}+a_{2}\to a_{1},\qquad a_{2}+a_{4}\to-a_{2},\qquad
a_{8}\to a_{3},&\nonumber\\
&a_{6}\to a_{4},\qquad a_{9}\to a_{5},\qquad a_{41}\to
a_{6},\qquad
a_{20}\to -\half\alpha\,a_{7},&\nonumber\\
&a_{43}\to a_{8},\qquad a_{47}\to a_{9},\qquad
a_{54}\to a_{10}\,\half\zeta,&\nonumber\\
&a_{55}+a_{56}\to a_{11}\,\rho,\qquad a_{57}\to
a_{12}\,\half\kappa,\qquad a_{59}\to a_{13}\,\half\beta,&
\end{eqnarray}
we get
\begin{eqnarray}
\Delta^{(-1)}&\!\!\!=\!\!\!&\Int\Bigl\{a_{1}\bigl(\Omega^{a}_{\mu}A^{a}_{\mu}
+g\e^{ab}\xi^{a}_{\mu}A^{b}_{\mu}c\bigr)
-a_{2}\xi^{a}_{\mu}D^{ab}_{\mu}c^{b}
+(a_{1}+a_{2}+a_{3})\bigl(\bar{N}^{a}_{\mu
i}D^{ab}_{\mu}\phi^{b}_{i}
-M^{a}_{\mu i}D^{ab}_{\mu}\bar{\omega}^{b}_{i}\bigr) \nonumber\\
&&+a_{3}\,L^{a}c^{a}
-(a_{3}+a_{4})\bar{\omega}^{a}_{i}\mathcal{M}^{ab}\phi^{b}_{i}
-a_{4}\,\bar{c}^{a}D^{ab}_{\mu}A^{b}_{\mu} +a_{5}\,Lc
+a_{6}\,\chi\,\bar{N}^{a}_{\mu i}M^{a}_{\mu i}
-\frac{\alpha}{2}\,a_{7}(i\bar{c}^{a}b^{a} -g\e^{ab}\bar{c}^{a}\bar{c}^{b}c)\nonumber\\
&&-\frac{\alpha}{2}(a_{7}-2a_{3})g^{2}\bar{\omega}^{a}_{i}\phi^{a}_{i}
(\bar{\phi}^{b}_{j}\phi^{b}_{j}
-\bar{\omega}^{b}_{j}\omega^{b}_{j}) +\alpha(
a_{7}-a_{3})g^{2}\bar{\omega}^{a}_{i}\phi^{a}_{i}\bar{c}^{b}c^{b}
+\half(a_{3}-a_{4}-\alpha a_{8})\lambda\, A^{a}_{\mu}A^{a}_{\mu}\nonumber\\
&&+\half a_{8}\,\tau A^{a}_{\mu}A^{a}_{\mu} +\alpha( a_{3}+a_{7}+
a_{9})\,\lambda\bigl( \bar{\phi}^{a}_{i}\phi^{a}_{i}
-\bar{\omega}^{a}_{i}\omega^{a}_{i}
-\bar{c}^{a}c^{a}\bigr)-(a_{3}+a_{9})\,\tau \bigl(
\bar{\phi}^{a}_{i}\phi^{a}_{i}
-\bar{\omega}^{a}_{i}\omega^{a}_{i}\bigr)
+a_{9}\,\tau\bar{c}^{a}c^{a}
\nonumber\\
&&-(a_{3}-a_{5})g\e^{ab}(\eta\bar{c}^{a}c^{b}
+\bar{\theta}_{i}\phi^{a}_{i}c^{b}
-\eta_{i}\bar{\omega}^{a}c^{b})+a_{10}\,\half\zeta\,J\lambda
+a_{11}\,\rho\,\s\lambda +a_{12}\,\half\kappa\,\s\tau
+a_{13}\,\half\beta\,\eta\theta\Bigr\}\;,
\end{eqnarray}
so that for the counterterm $\Sigma_{\mathrm{CT}}$ we get
\begin{eqnarray}
\Sigma_{\mathrm{CT}}&\!\!\!=\!\!\!&\Int\biggl\{
(a_{0}+2a_{1})\biggl[\frac{1}{2}(\partial_{\mu}A^{a}_{\nu})
(\partial_{\mu}A^{a}_{\nu}-\partial_{\nu}A^{a}_{\mu})-
g\varepsilon^{ab}(\partial_{\mu}A^{a}_{\nu})(A_{\mu}A^{b}_{\nu}
-A_{\nu}A^{b}_{\mu})
+g\varepsilon^{ab}(\partial_{\mu}A_{\nu})A^{a}_{\mu}A^{b}_{\nu}\nonumber\\
&&+\frac{g^{2}}{2}(A_{\mu}A_{\mu}A^{a}_{\nu}A^{a}_{\nu}+
A_{\mu}A_{\nu}A^{a}_{\mu}A^{a}_{\nu})\biggr]+
\frac{a_{0}}{2}\,(\partial_{\mu}A_{\nu})(\partial_{\mu}A_{\nu}
-\partial_{\nu}A_{\mu})+
(a_{0}+4a_{1})\frac{g^{2}}{4}A^{a}_{\mu}A^{a}_{\mu}A^{b}_{\nu}A^{b}_{\nu}\nonumber\\
&&+ i(a_{1}-a_{4})b^{a}D^{ab}_{\mu}A^{b}_{\mu}
-(a_{3}+a_{4})\Bigl[ \bar c^{a}\partial^{2}c^{a}-\bar
c^{a}g\varepsilon^{ab}(\partial_{\mu}A_{\mu})c^{b} -2\bar
c^{a}g\varepsilon^{ab}A_{\mu}\partial_{\mu}c^{b}
-g^{2}\bar c^{a}c^{a}A_{\mu}A_{\mu}\Bigr]\nonumber\\
&&+(a_{3}+a_{4})\Bigl[ \bar
\phi^{a}_{i}\partial^{2}\phi^{a}_{i}-\bar
\phi^{a}_{i}g\varepsilon^{ab}(\partial_{\mu}A_{\mu})\phi^{b}_{i}
-2\bar
\phi^{a}_{i}g\varepsilon^{ab}A_{\mu}\partial_{\mu}\phi^{b}_{i}
-g^{2}\bar \phi^{a}_{i}\phi^{a}_{i}A_{\mu}A_{\mu}\Bigr]\nonumber\\
&&-(a_{3}+a_{4})\Bigl[ \bar
\omega^{a}_{i}\partial^{2}\omega^{a}_{i}-\bar
\omega^{a}_{i}g\varepsilon^{ab}(\partial_{\mu}A_{\mu})\omega^{b}_{i}
-2\bar
\omega^{a}_{i}g\varepsilon^{ab}A_{\mu}\partial_{\mu}\omega^{b}_{i}
-g^{2}\bar \omega^{a}_{i}\omega^{a}_{i}A_{\mu}A_{\mu}\Bigr]\nonumber\\
&&+(2a_{1}-a_{3}-a_{4})g^{2}\varepsilon^{ac}\varepsilon^{bd} (\bar
c^{a}c^{b}-\bar \phi^{a}_{i}\phi^{b}_{i} +\bar
\omega^{a}_{i}\omega^{b}_{i})A^{c}_{\mu}A^{d}_{\mu}
+(a_{1}-a_{4}-a_{5})g\varepsilon^{ab}\bar
c^{a}cD^{bc}_{\mu}A^{c}_{\mu}
\nonumber\\
&& -(a_{1}-a_{3})(\Omega_{\mu}+\partial_{\mu}\bar
c)g\varepsilon^{ab} A^{a}_{\mu}c^{b}
+(a_{1}-2a_{3}-a_{4})\Bigl[2g^{2}\varepsilon^{ab}\varepsilon^{cd}
\bar\omega^{a}_{i}A^{c}_{\mu}c^{d}\partial_{\mu}\phi^{b}_{i}
+g^{2}\varepsilon^{ab}\varepsilon^{cd}
\bar\omega^{a}_{i}\partial(A^{c}_{\mu}c^{d})\phi^{b}_{i}
\nonumber\\
&&+2g^{3}\varepsilon^{bc}\bar\omega^{a}_{i}\phi^{a}_{i}
A_{\mu}A^{b}_{\mu}c^{c}-g^{2}(\varepsilon^{ac}\varepsilon^{bd}+
\varepsilon^{ad}\varepsilon^{bc})\bar\omega^{a}_{i}A^{d}_{\mu}
(D^{ce}_{\mu}c^{e})\phi^{b}_{i}\Bigr]
+(2a_{1}-a_{3}-a_{4}-a_{5})g^{3}(\delta^{ae}\varepsilon^{bd}+
\delta^{be}\varepsilon^{ad})\nonumber\\
&&\times\bar\omega^{a}_{i} A^{d}_{\mu}A^{e}_{\mu}\phi^{b}_{i}c
-(a_{2}+2a_{3})g^{2}\varepsilon^{ab}\varepsilon^{cd}
(\bar{N}^{a}_{\mu i}\phi^{b}_{i}+M^{a}_{\mu i}\bar\omega^{b}_{i})
A^{c}_{\mu}c^{d}+(a_{1}+a_{2}+a_{3})\Omega^{a}_{\mu}D^{ab}_{\mu}c^{b}
-a_{2}K^{a}_{\mu}D^{ab}_{\mu}c^{b}
\nonumber\\
&&+a_{5}g\varepsilon^{ab}
K^{a}_{\mu}A^{b}_{\mu}c-(a_{1}+a_{2}+a_{3}+a_{5})g\varepsilon^{ab}
\xi^{a}_{\mu}(D^{bc}_{\mu}c^{c})c
+(a_{2}+2a_{3})\frac{g^{2}}{2}\varepsilon^{ab}\varepsilon^{cd}
\xi^{a}_{\mu}A^{b}_{\mu}c^{c}c^{d}-a_{5}g\varepsilon^{ab}L^{a}c^{b}c
\nonumber\\
&&-(2a_{3}-a_{5})\frac{g^{2}}{2}\varepsilon^{ab}Lc^{a}c^{b}
+a_{5}(\Omega_{\mu}+\partial_{\mu}\bar c)\partial_{\mu}c
+a_{3}g^{2}\varepsilon^{ab}\varepsilon^{cd}(\bar{X}^{a}_{i}\phi^{b}_{i}-
Y^{a}_{i}\bar\omega^{b}_{i})c^{c}c^{d}-a_{5}g\varepsilon^{ab}
(\bar{Y}^{a}_{i}\phi^{b}_{i}-\bar{X}^{a}_{i}\omega^{b}_{i}
\nonumber\\
&&+X^{a}_{i}\bar\omega^{b}_{i}-Y^{a}_{i}\bar\phi^{b}_{i})c-
(a_{3}+a_{4}+a_{5})\Bigl[2g\varepsilon^{ab}\bar\omega^{a}_{i}
(\partial_{\mu}c)\partial_{\mu}\phi^{b}_{i}+
g\varepsilon^{ab}\bar\omega^{a}_{i}(\partial^{2}c)\phi^{b}_{i}
+2g^{2}\bar\omega^{a}_{i}\phi^{a}_{i}A_{\mu}\partial_{\mu}c\Bigr]
\nonumber\cr
&&+(a_{1}+a_{2}+a_{3}+a_{5})g\varepsilon^{ab}(\partial_{\mu}c)
(\bar{N}^{a}_{\mu i}\phi^{b}_{i}-M^{a}_{\mu i}\bar\omega^{b}_{i})
-(a_{1}+a_{2}+a_{3})(\bar{M}^{a}_{\mu i}D^{ab}_{\mu}\phi^{b}_{i}+
\bar{N}^{a}_{\mu i}D^{ab}_{\mu}\omega^{b}_{i}+ M^{a}_{\mu
i}D^{ab}_{\mu}\bar\phi^{b}_{i}\nonumber\\
&&+ N^{a}_{\mu i}D^{ab}_{\mu}\bar\omega^{b}_{i})
-a_{6}\,\chi(\bar{M}^{a}_{\mu i}M^{a}_{\mu i}+ \bar{N}^{a}_{\mu
i}N^{a}_{\mu i}) +\frac{\alpha}{2}\Bigl[a_{7}b^{a}b^{a}
+2i(a_{7}-a_{5})g\e^{ab}b^{a}\bar{c}^{b}c
-g^{2}(a_{7}-2a_{3})\bigl(\bar{\phi}^{a}_{i}\phi^{a}_{i}\nonumber\\
&&-\bar{\omega}^{a}_{i}\omega^{a}_{i}-\bar{c}^{a}c^{a}\bigr)
\bigl(\bar{\phi}^{b}_{j}\phi^{b}_{j}
-\bar{\omega}^{b}_{j}\omega^{b}_{j}-\bar{c}^{b}c^{b}\bigr)
-2ig^{2}(a_{7}-2a_{3})\bar{\omega}^{a}_{i}\phi^{a}_{i}b^{b}c^{b}
+2g^{3}(a_{7}-2a_{3}-a_{5})\bar{\omega}^{a}_{i}\phi^{a}_{i}
\e^{bc}\bar{c}^{b}c^{c}c\Bigr]\nonumber\\
&&+(2a_{1}+a_{3}-a_{4}-\alpha a_{8})\half JA^{a}_{\mu}A^{a}_{\mu}
+(a_{1}-a_{4}-\alpha a_{8})\lambda A^{a}_{\mu}D^{ab}_{\mu}c^{b}
+(a_{3}+a_{9})\bigl(\s\bar{c}^{a}c^{a} -i\tau
b^{a}c^{a}\bigr)\nonumber\\
&&+(a_{3}+a_{5}+a_{9})\tau g\e^{ab}\bar{c}^{a}c^{b}c
-(2a_{3}-a_{5})g\e^{ab}\bigl(\theta\bar{c}^{a}c^{b}
+\bar{\eta}_{i}\phi^{a}_{i}c^{b} +\eta_{i}\bar{\phi}^{a}_{i}c^{b}
+\bar{\theta}_{i}\omega^{a}_{i}c^{b}
+\theta_{i}\bar{\omega}^{a}_{i}c^{b} -i\eta
b^{a}c^{b}\bigr)\nonumber\\
&&+2g^{2}a_{3}\bigl(\eta\bar{c}^{a}c^{a}c
+\bar{\theta}_{i}\phi^{a}_{i}c^{a}c
-\eta_{i}\bar{\omega}^{a}_{i}c^{a}c\bigr) +a_{8}\half\sigma
A^{a}_{\mu}A^{a}_{\mu} +a_{8}\tau A^{a}_{\mu}D^{ab}_{\mu}c^{b}
+\alpha(a_{3}+a_{7}+a_{9})
\Bigl[J\bigl(\bar{\phi}^{a}_{i}\phi^{a}_{i}\nonumber\\
&&-\bar{\omega}^{a}_{i}\omega^{a}_{i} -\bar{c}^{a}c^{a}\bigr)
+\lambda(ib^{a}c^{a} -g\e^{ab}\bar{c}^{a}c^{b}c)\Bigr]
-(a_{3}+a_{9})\s (\bar{\phi}^{a}_{i}\phi^{a}_{i}
-\bar{\omega}^{a}_{i}\omega^{a}_{i}\bigr)
+a_{10}\frac{\zeta}{2}\,J^{2} +a_{11}\,\rho\,J\s\nonumber\\
&&+a_{12}\frac{\kappa}{2}\,\s^{2}
+a_{13}\frac{\beta}{2}\,\theta^{2}\biggr\}\;. \label{CT}
\end{eqnarray}
After the characterization of the most general local counterterm
$\S_{\mathrm{CT}}$ compatible with all constraints,
eqs.\eqref{constraints}, we still have to check if it can be
reabsorbed through a multiplicative redefinition of the fields,
sources and parameters of the starting action $\S$, according to
\begin{equation}
\S[\Psi_{0},\psi_{0},\mathcal{J}_{0},\vartheta_{0},
\Omega_{0},K_{0},\lambda_{0},\tau_{0},J_{0},\s_{0}]=
\S[\Psi,\psi,\mathcal{J},\vartheta, \Omega,K,\lambda,\tau,J,\s]
+\epsilon\,\S_{\mathrm{CT}}+O(\epsilon^{2})\;,
\end{equation}
where
\begin{equation}
\Psi_{0}=\widetilde{Z}^{1/2}_{\Psi}\,\Psi\;,\qquad
\psi_{0}=Z^{1/2}_{\psi}\,\psi\;,\qquad
\mathcal{J}_{0}=Z_{\mathcal{J}}\,\mathcal{J},\qquad
\vartheta_{0}=Z_{\vartheta}\,\vartheta,
\end{equation}
with
\begin{eqnarray}
\Psi&\equiv& \{A^{a}_{\mu},b^{a},c^{a},\bar{c}^{a}\}\;,\nonumber\\
\psi&\equiv&\{A_{\mu}, b, c, \bar{c}, \phi^{a}_{i},
\bar{\phi}^{a}_{i},
\omega^{a}_{i},\bar{\omega}^{a}_{i}\}\;,\nonumber\\
\mathcal{J}&\equiv&\{\xi^{a}_{\mu}, L^{a}, L, \Omega_{\mu},
X^{a}_{i}, \bar{X}^{a}_{i}, Y^{a}_{i}, \bar{Y}^{a}_{i}, M^{a}_{\mu
i}, \bar{M}^{a}_{\mu i}, N^{a}_{\mu i}, \bar{N}^{a}_{\mu i}, \eta,
\theta, \bar{\eta}_{i}, \eta_{i}, \bar{\theta}_{i},
\theta_{i}\}\;,\nonumber\\
\vartheta&\equiv&\{g, \alpha, \chi, \zeta, \rho, \kappa,
\beta\}\;.
\end{eqnarray}
Moreover, by taking into account the mixing of the sources
displaying the same quantum numbers, \textit{i.e},
$(\Omega^{a}_{\mu},K^{a}_{\mu})$, $(\lambda,\tau)$ and $(J,\s)$,
we shall set
\begin{equation}
\left(\begin{matrix} \Omega_{0\mu}^{\phantom{0}a}\\
K^{\phantom{0}a}_{0\mu}
\end{matrix}\right)=\mathbb{Z}_{\Omega K}
\left(\begin{matrix}
\Omega_{\mu}^{a}\\
K^{a}_{\mu}
\end{matrix}\right),\qquad
\left(\begin{matrix}
\lambda_{0}\\
\tau_{0}
\end{matrix}\right)=\mathbb{Z}_{\lambda\tau}
\left(\begin{matrix}
\lambda\\
\tau
\end{matrix}\right),\qquad
\left(\begin{matrix}
J_{0}\\
\s_{0}
\end{matrix}\right)=\mathbb{Z}_{J\s}
\left(\begin{matrix}
J\\
\s
\end{matrix}\right),
\end{equation}
where the $\mathbb{Z}$-matrices are given by
\begin{equation}
\mathbb{Z}_{\Omega K}= \mathbb{I}+\epsilon \left(\begin{matrix}
z_{\Omega}&z_{\Omega K}\\
z_{K\Omega}&z_{K}
\end{matrix}\right),\qquad
\mathbb{Z}_{\lambda\tau}= \mathbb{I}+\epsilon \left(\begin{matrix}
z_{\lambda}&z_{\lambda\tau}\\
z_{\tau\lambda}&z_{\tau}
\end{matrix}\right),\qquad
\mathbb{Z}_{J\s}= \mathbb{I}+\epsilon \left(\begin{matrix}
z_{J}&z_{J\s}\\
z_{\s J}&z_{\s}
\end{matrix}\right).\qquad
\end{equation}
By direct inspection of $\S_{\mathrm{CT}}$, the renormalization
factors are found to be
\begin{eqnarray}
&Z^{1/2}_{A}=Z^{-1}_{g},\qquad
\widetilde{Z}^{1/2}_{\bar{c}}=\widetilde{Z}^{1/2}_{c},&\nonumber\\
&\widetilde{Z}^{1/2}_{b}=Z_{g}Z^{-1/2}_{\bar{c}}\widetilde{Z}^{1/2}_{c},\qquad
Z_{b}^{1/2}=Z_{g},&\nonumber\\
&Z_{\phi}^{1/2}=Z_{\bar{\phi}}^{1/2}=\widetilde{Z}^{1/2}_{c},&\nonumber\\
&Z^{1/2}_{\bar\omega}=Z^{-1}_{g}Z^{1/2}_{\bar{c}}\widetilde{Z}^{1/2}_{c},\qquad
Z^{1/2}_{\omega}=Z_{g}Z^{-1/2}_{\bar{c}}\widetilde{Z}^{1/2}_{c},&\nonumber\\
&Z_{\bar{M}}=Z_{M},\qquad
Z_{N}=Z_{g}Z^{-1/2}_{\bar{c}}Z_{M},\qquad
Z_{\bar{N}}=Z_{g}^{-1}Z^{1/2}_{\bar{c}}Z_{M},&\nonumber\\
&Z_{X}=\widetilde{Z}^{-1/2}_{c},\qquad
Z_{\bar{X}}=Z^{-2}_{g}Z_{\bar{c}}\widetilde{Z}^{-1/2}_{c},&\nonumber\\
&Z_{Y}=Z_{\bar{Y}}=Z^{-1}_{g}Z^{1/2}_{\bar{c}}\widetilde{Z}^{-1/2}_{c},&\nonumber\\
&Z_{\theta}=Z_{\bar\eta_{i}}=Z_{\eta_{i}}=
Z_{g}^{-1}Z_{\bar{c}}^{1/2} Z_{c}^{-1/2},&\nonumber\\
&Z_{\eta}=Z_{\bar\theta_{i}}=Z^{-2}_{g}Z_{\bar{c}}Z_{c}^{1/2},
\qquad Z_{\theta_{i}}=Z_{c}^{-1/2},&\nonumber\\
&\widetilde{Z}_{L}=Z^{-1}_{g}Z^{1/2}_{\bar{c}}\widetilde{Z}^{-1/2}_{c},\qquad
{Z}_{L}=Z^{-1}_{g}Z^{1/2}_{\bar{c}}{Z}^{-1/2}_{c},&\nonumber\\
&Z_{\xi}=Z_{M},\qquad Z_{\Omega}=Z^{1/2}_{\bar{c}},&
\end{eqnarray}
with
\begin{eqnarray}
\widetilde{Z}^{1/2}_{A}&=&1+\epsilon\Bigl(\frac{a_{0}}{2}+a_{1}\Bigr)\;,\qquad
Z_{g}\;\;=\;\;1-\epsilon\,\frac{a_{0}}{2}\;,\nonumber\\
\widetilde{Z}^{1/2}_{c}&=&1-\epsilon\,\frac{a_{3}+a_{4}}{2}\;,\qquad
Z^{1/2}_{\bar{c}}\;\;=\;\;1-\epsilon\,\frac{a_{3}-a_{4}}{2}\;,\nonumber\\
Z^{1/2}_{c}&=&1+\epsilon\Bigl(\frac{a_{3}-a_{4}}{2}-a_{5}\Bigr)\;,\nonumber\\
Z_{M}&=&1-\epsilon\Bigl(a_{1}+a_{2}+\frac{a_{3}-a_{4}}{2}\Bigr)\;,\nonumber\\
Z_{\chi}&=&1+\epsilon\bigl(2a_{1}+2a_{2}+a_{3}-a_{4}-a_{6}\bigr)\;,\nonumber\\
Z_{\alpha}&=&1+\epsilon\bigl(a_{0}+2a_{4}+a_{7}\bigr)\;,\nonumber\\
Z_{\zeta}&=&1+\epsilon\Bigl[2a_{0}-2\Bigl(1+\alpha\frac{\rho}{\zeta}\Bigr)a_{3}
+2a_{4} - 2\alpha\frac{\rho}{\zeta}a_{7} + 2\alpha a_{8}
-2\alpha\frac{\rho}{\zeta}a_{9}+a_{10}\Bigr]\;,\nonumber\\
Z_{\rho}&=&1+\epsilon\Bigl[a_{0}-(1+\alpha)a_{3}+a_{4} -\alpha
a_{7}
-\Bigl(-\alpha+\frac{\zeta}{\rho}\Bigr)(a_{8}-a_{9})+a_{11}\Bigr]\;,\nonumber\\
Z_{\kappa}&=&1-\epsilon
\Bigl[2a_{4}+2\frac{\rho}{\kappa}a_{8}-2a_{9}-a_{12}\Bigr]\;,\nonumber\\
Z_{\beta}&=&1-\epsilon\bigl(a_{0}-2a_{3}+2a_{4}+2a_{5}-a_{13}\bigr)\;,
\end{eqnarray}
and
\begin{eqnarray}
\mathbb{Z}_{\Omega K}&=&\mathbb{I}+\epsilon \left(\,\,
\begin{matrix}
-a_{1}-a_{2}-\half(a_{3}-a_{4})&\vline&a_{2}\!\phantom{\Bigl|}\\\hline
0&\vline&-a_{1}-\half(a_{3}-a_{4})\!\phantom{\Bigl|}
\end{matrix}
\,\,\right)\;,\nonumber\\
\mathbb{Z}_{\lambda\tau}&=&\mathbb{I}+\epsilon \left(\,\,
\begin{matrix}
-\half(a_{0}-a_{3}+a_{4})-\alpha
a_{8}&\vline&a_{8}\!\phantom{\Bigl|}\\\hline
+\alpha(a_{3}+a_{7}+a_{9})
&\vline&\half(a_{0}-a_{3}+3a_{4})-a_{9}\!\phantom{\Bigl|}
\end{matrix}
\,\,\right)\;,\nonumber\\
\mathbb{Z}_{J\s}&=&\mathbb{I}+\epsilon \left(\,\,
\begin{matrix}
-a_{0}+a_{3}-a_{4}-\alpha
a_{8}&\vline&a_{8}\!\phantom{\Bigl|}\\\hline
+\alpha(a_{3}+a_{7}+a_{9})& \vline&a_{4}-a_{9}\!\phantom{\Bigl|}
\end{matrix}
\,\,\right)\;.
\end{eqnarray}
This concludes the proof of the renormalizability of the complete
calssical starting action $\S$.
\section{Conclusion}
In this paper the gluon and ghost propagators have been
investigated by taking into account the effects of the Gribov
copies as well as of dimension two operators. The output of our
results is summarized in Sect.2, where the expressions for the
tree level propagators can be found, being in good agreement with
the most recent lattice data \cite{Mendes:2006kc}. \\\\Certainly,
much work is needed in order to reach a better understanding of
the maximal Abelian gauge. Nevertheless, the results which we have
obtained enable us to strengthen the fact that the agreement with
the lattice data has been obtained only when the effects of the
Gribov copies and of the dimension two operators have been
simultaneously encoded in the starting Lagrangian, which enjoys
the important property of being renormalizable. This point can be
better clarified by the following considerations:
\begin{itemize}
\item {\bf The quantization procedure and the issue of the Gribov
copies} \\
The starting point to analyze Yang-Mills theories at the quantum
level is by means of the Faddeev-Popov quantization formula, based
on the introduction of a gauge fixing and of the corresponding
ghost term. It is known that such a procedure is plagued by the
existence of the Gribov copies. A full  resolution of this issue,
amounting to restrict the domain of integration in the Feynman
path integral to the fundamental modular region, is still
unavailable. A partial solution to this problem consists of
restricting the domain of integration to the Gribov region $
\Omega$, which is still affected by Gribov copies. Although this
procedure does not eliminate all copies, it has the advantage of
being effectively implementable. As we learn from the work of
Zwanziger \cite{Zwanziger:1989mf,Zwanziger:1992qr} in the Landau
gauge, the restriction to the region $\Omega$ is achieved through
the introduction in the Yang-Mills action of a nonlocal operator,
known as the horizon function. This nonlocal operator can be cast
in local form by introducing a set of additional localizing
fields. Remarkably, the resulting local action turns out to be
renormalizable \cite{Zwanziger:1989mf,Zwanziger:1992qr}. This
procedure has been successfully adapted to the maximal Abelian
gauge \cite{Capri:2005tj,Capri:2006cz}. A second point to be
noticed is that the introduction of the horizon function in its
local form is equivalent to the introduction of a specific
dimension two operator. In fact, the Gribov-Zwanziger gap equation
\cite{Gribov:1977wm,Zwanziger:1989mf,Zwanziger:1992qr} determining
the Gribov parameter $\gamma$, namely
\begin{equation}
\frac{\delta \Gamma}{\delta\gamma}=0 \; , \label{gapeq}
\end{equation} with $\Gamma$ being the 1PI effective action, is
equivalent to require the existence of a nonvanishing dimension
two condensate. In the case of the Landau gauge, this condensate
is given by \cite{Zwanziger:1989mf,Zwanziger:1992qr}
\begin{equation}
\langle f^{ABC}A^{A}_{\mu}(x)(\phi^{BC}_\mu(x) -
{\bar\phi}^{BC}_{\mu}(x))\rangle \neq 0 \;, \label{landauc}
\end{equation}
where $\phi^{BC}_\mu, {\bar\phi}^{BC}_{\mu}$ stand for the
localizing fields and the indices $A,B,C$ belong to the adjoint
representation of $SU(2)$. In the case of the maximal Abelian
gauge the corresponding condensate is given by $\langle
\e^{ab}A_{\mu}(x)(\phi^{ab}_\mu(x) -
{\bar\phi}^{ab}_{\mu}(x)\rangle$. The same feature holds in the
Coulomb gauge, see \cite{Zwanziger:2007zz} for a review.  \item
{\bf Introduction of the dimension two operators} \\As mentioned
before, the inclusion of the horizon function is equivalent to the
introduction of a dimension two field operator in the localizing
fields. Therefore, we can look for other dimension two operators
which can be added to the theory, provided one is able to maintain
renormalizability. From this point of view, the introduction of
the three dimension two operators
$\mathcal{O}_{A^{2}}=\half\,A^{a}_{\mu}A^{a}_{\mu}$,
$\mathcal{O}_{\bar{f}f}= (\bar{\phi}^{b}_{i}\phi^{b}_{i}
-\bar{\omega}^{b}_{i}\omega^{b}_{i} -\bar{c}^{b}c^{b})$,
$\mathcal{O}_{\mathrm{ghost}}=(\e^{ab}\bar{c}^{a}c^{b})$ looks
very natural. It is remarkable that these three operators can be
simultaneously added to the horizon term in a way which preserves
renormalizability. We also notice that all three operators
considered here have their analogue in the Landau gauge, see
\cite{Dudal:2003pe,Dudal:2007cw} and refs. therein. In much the
same way as the horizon function, these operators carry
nonperturbative information, encoded in the corresponding
condensates. \\\\The good agreement of our results with the
lattice data can be taken as evidence of the fact these dimension
two operators play a relevant role in the infrared. For example,
without the introduction of the two operators
$\mathcal{O}_{\bar{f}f}$, $\mathcal{O}_{\mathrm{ghost}}$, the
infrared behavior of the off-diagonal ghost propagator would be
deeply different from that of eq.\eqref{symmghost}. Instead, it
would have displayed an enhanced behavior of the type  $1/k^4$, as
reported in our previous investigation \cite{Capri:2005tj}, where
only the horizon function and the gluon condensate
$\mathcal{O}_{A^{2}}$ were taken into account. The same occurs for
the diagonal gluon propagator, eq.\eqref{diaggluon}. Without the
introduction of $\mathcal{O}_{\bar{f}f}$ it would be vanishing at
$k=0$.
\\\\We remark that the same features have been detected in
the Landau gauge, where the most recent lattice data
\cite{Cucchieri:2007md,Bogolubsky:2007ud,Cucchieri:2007rg} point
towards a finite and nonvanishing gluon propagator at $k=0$, while
exhibiting a less enhanced ghost propagator. As discussed in
\cite{Dudal:2007cw} these features can be accounted for by
considering the effects of dimension two operators, which nicely
fit within the Gribov-Zwanziger framework.

\end{itemize}

\section*{Acknowledgments}

S. P. Sorella thanks D. Dudal, A. Cucchieri, T. Mendes, N.
Vandersickel and D. Zwanziger for useful discussion. The Conselho
Nacional de Desenvolvimento Cient\'{i}fico e Tecnol\'{o}gico
(CNPq-Brazil), the Faperj, Funda{\c{c}}{\~{a}}o de Amparo {\`{a}}
Pesquisa do Estado do Rio de Janeiro, the SR2-UERJ and the
Coordena{\c{c}}{\~{a}}o de Aperfei{\c{c}}oamento de Pessoal de
N{\'{i}}vel Superior (CAPES) are gratefully acknowledged for
financial support.

\end{document}